\documentstyle[12pt]{article}

\advance \topmargin by -\headheight
\advance \topmargin by -\headsep

\evensidemargin \oddsidemargin
\def\ie{{\em i.e.}}
\def\ie{\hbox{\it i.e.}}

\def\CC{{\mathchoice
{\rm C\mkern-8mu\vrule height1.45ex depth-.05ex
width.05em\mkern9mu\kern-.05em}
{\rm C\mkern-8mu\vrule height1.45ex depth-.05ex
width.05em\mkern9mu\kern-.05em}
{\rm C\mkern-8mu\vrule height1ex depth-.07ex
width.035em\mkern9mu\kern-.035em}
{\rm C\mkern-8mu\vrule height.65ex depth-.1ex
width.025em\mkern8mu\kern-.025em}}}

\def\RR{{\rm I\kern-1.6pt {\rm R}}}

\def\ZZ{{\rm Z}\kern-3.8pt {\rm Z} \kern2pt}
\def\II{\relax{\rm I\kern-.18em I}}
\def\IP{\relax{\rm I\kern-.18em P}}

\def\np{Nucl. Phys.}
\def\pl{Phys. Lett.}

\def\jhep{J. High Energy Phys.}

\newcommand{\beq}{\begin{equation}}
\newcommand{\eeq}{\end{equation}}
\newcommand{\rc}{\nonumber\\}
\newcommand{\bear}{\begin{eqnarray}}
\newcommand{\eear}{\end{eqnarray}}

\newfont{\namefont}{cmr10}
\newfont{\addfont}{cmti7 scaled 1440}
\newfont{\boldmathfont}{cmbx10}
\newfont{\headfontb}{cmbx10 scaled 1728}
\renewcommand{\theequation}{{\rm\thesection.\arabic{equation}}}
\begin{document}
\begin{titlepage}

\begin{center} \Large \bf Wrapped branes with fluxes in 8d gauged supergravity

\end{center}

\vskip 0.3truein
\begin{center}
J. D. Edelstein${}^{\,\sharp\,\dagger}$
\footnote{edels@fisica.unlp.edu.ar,~jedels@math.ist.utl.pt},
A. Paredes${}^{\,*}$
\footnote{angel@fpaxp1.usc.es}
and
A.V. Ramallo${}^{\,*}$
\footnote{alfonso@fpaxp1.usc.es}

\vspace{0.3in}
${}^{\sharp\,}$Departamento de Matem\'atica, Instituto Superior Tecnico \\
Av. Rovisco Pais, 1049--001, Lisboa, Portugal

\vspace{0.3in}

${}^{\dagger\,}$Instituto de F\'\i sica de La Plata -- Conicet, Universidad
Nacional de La Plata \\
C.C. 67, (1900) La Plata, Argentina

\vspace{0.3in}

${}^{\,*}$Departamento de F\'\i sica de Part\'\i culas, Universidad de Santiago
de Compostela \\
E-15782 Santiago de Compostela, Spain
\vspace{0.3in}

\end{center}
\vskip 1truein

\begin{center}
\bf ABSTRACT
\end{center}

We study the gravity dual of several wrapped D--brane configurations in
presence of 4-form RR fluxes partially piercing the unwrapped directions.
We present a systematic approach to obtain these solutions from those without
fluxes. We use D=8 gauged supergravity as a starting point to build up
these solutions. The configurations include (smeared) M2--branes at the
tip of a $G_2$ cone on $S^3 \times S^3$, D2--D6 branes with the latter
wrapping a special Lagrangian 3-cycle of the complex deformed conifold
and an holomorphic sphere in its cotangent bundle $T^*S^2$, D3--branes
at the tip of the generalized resolved conifold, and others obtained by
means of T duality and KK reduction. We elaborate on the corresponding
${\cal N}=1$ and ${\cal N}=2$ field theories in $2+1$ dimensions.

\vskip2.6truecm
\leftline{US-FT-2/02\hfill July 2002}
\leftline{hep-th/0207127}
\smallskip
\end{titlepage}
\setcounter{footnote}{0}


\setcounter{equation}{0}
\section{Introduction}
\medskip

The low energy dynamics of a collection of D--branes wrapping supersymmetric
cycles is governed, when the size of the cycle is taken to zero, by a lower
dimensional supersymmetric gauge theory with less than sixteen supercharges.
The non-trivial geometry of the world--volume leads to a gauge
theory in which supersymmetry is appropriately twisted \cite{bvs}, the
amount of preserved supersymmetries having to do with the way in which the
cycle is embedded in a higher dimensional space. In a generalization of the
AdS/CFT correspondence \cite{jm}, when the number of branes is taken to be
large, the near horizon limit of the corresponding supergravity solutions
provide gravity duals to the field theories arising on the world--volume of
the D--branes \cite{mn}. The gravitational description of the strong
coupling regime of these gauge theories allows for a geometrical approach
to the study of such important aspects of their infrared dynamics as, for
example, chiral symmetry breaking, gaugino condensation, confinement and
the existence of a mass gap \cite{mn,ks}.

An exhaustive study of the gravity/gauge theory correspondence for flat
D--branes --whose low energy dynamics is dictated, in general, by
non--conformal field theories-- as well as the intricate phase structure
of their RG flows, was undertaken in ref.\cite{imsy}. In the case of
theories with less than sixteen supercharges, the fact that there are too
many possibilities in choosing the D--branes, the cycles, and the manifolds
embedding them, prevent the very existence of an analogous comprehensive
work. Yet, many cases have been considered so far \cite{gauged}--\cite{gkmw}.
A natural framework to perform the above mentioned twisting is given by
lower dimensional gauged supergravities. Their solutions usually correspond
to the near horizon limit of D--brane configurations thus giving directly
the gravity dual description of the gauge theories living on their
world--volumes. This approach, started in \cite{mn}, has been widely
followed throughout the literature on the subject.

Gauged supergravities have several forms coming from the dimensional
reduction of the highest dimensional supergravities \cite{ssch}. Turning
them on amounts to the introduction of other branes into the system in the
form of either localized or smeared intersections and overlappings. Many
of these configurations correspond to extremely interesting supersymmetric
gauge theories. In particular, these configurations give rise to a
world--volume dynamics whose description, at different energy scales, is
given by increasingly richer phases connected by RG flows. See, for example,
\cite{ps}--\cite{gns}.

The purpose of this article is to study the effect of turning on 4-form
fluxes in the non compact directions of D6--branes wrapping supersymmetric
cycles. We shall analyze different configurations that correspond to
(smeared) M2--branes at the tip of a $G_2$ cone on $S^3 \times S^3$,
D2--D6 branes with the latter wrapping a special Lagrangian 3-cycle of
the complex deformed conifold and an holomorphic sphere in its cotangent
bundle $T^*S^2$, D3--branes at the tip of the generalized resolved
conifold, find their supergravity duals and explore their T duals.
Some of the ten dimensional solutions display the phenomenon of
supersymmetry without supersymmetry \cite{dlp}.
From the eleventh dimensional point of view, all the solutions amount to
deformations of the purely gravitational backgrounds found in \cite{en}
corresponding to the small resolution of the conifold and a manifold of
$G_2$ holonomy which is topologically $\RR^4\times S^3$. We shall
construct these supergravity solutions by implementing the required
topological twisting in maximal eight dimensional gauged supergravity
\cite{ss}. We further elaborate on the corresponding ${\cal N}=1$ and
${\cal N}=2$ field theories in $2+1$ dimensions.

The plan for the rest of the paper is as follows. In section $2$ we treat
the case of D6--branes wrapping a special Lagrangian submanifold of the
complex deformed conifold $T^*S^3$ in presence of 4-form fluxes. We derive
the BPS equations both through the vanishing condition for supersymmetry
transformations of the fermions as well as from the domain wall equations
resulting from the effective Routhian obtained by inserting the ansatz into
the 8d gauged supergravity Lagrangian. We obtain the general solution and
uplift it to eleven dimensions. It describes a configuration of smeared
M2--branes transverse to a (resolved) $G_2$ cone on $S^3 \times S^3$.
We then discuss all possible reductions to D=10, their T duals, and the
corresponding ${\cal N}=1$ dual field theories in $2+1$ dimensions.
In section $3$ we include 4-form fluxes in a configuration of D6 branes
wrapping a holomorphic $S^2$ in the cotangent bundle $T^*S^2$. We
proceed following the same avenue of the previous section, and find a
configuration of smeared M2--branes transverse to the generalized
resolved conifold. Again, we study the reductions to D=10, their T
duals, and the ${\cal N}=2$ dual field theories arising in $2+1$
dimensions. We present non--supersymmetric supergravity solutions
obtained by these means that correspond to smeared D2--D6 wrapping
a two--cycle and also to D4 configurations. These are examples of the
above referred supersymmetry without supersymmetry behavior.

We discuss the results and present our conclusions in section 4.
In appendix A we have collected the lagrangian and equations of motion
of D=8 gauged supergravity, together with the corresponding uplifting
formulae for the metric and the forms. We explain, in appendix B, how
to find an effective lagrangian for a given ansatz for the eight
dimensional fields when the 4-form $G$ is turned on. This involves a
subtle sign flip when constructing the Routhian after integrating out
the corresponding potential. Finally, we show in appendix C that the
deformation of the background produced by the inclusion of a 4-form
amounts to the appearance of warp factors. This result generalizes a
similar one recently obtained by Hern\'andez and Sfetsos in the study of
branes with fluxes wrapping spheres \cite{hs2}.

\setcounter{equation}{0}
\section{D6-branes wrapped on a 3-sphere with 4-form}
\medskip

In this section we will obtain supergravity solutions which correspond to
D6-branes wrapped on a three sphere with a flux turned on its worldvolume. We
will derive these solutions first in eight-dimensional gauged supergravity and
then we will uplift the result to eleven and ten dimensions.
The bosonic truncation of eight dimensional supergravity
relevant for our purposes contains the metric $g_{\mu\nu}$, some scalar
fields, an $SU(2)$ gauge potential $A^i$ and a three-form potential (whose
field strength we will denote by $G$). Notice that, in general, this is, an
inconsistent truncation of Salam and Sezgin's theory: $G$ acts as a
non--linear source for some of the forms we have turned off. However,
we will consider solutions that are fully compatible with the equations
of motion of D=8 gauged supergravity. To this end, $G$ must obey the
following constraints:
\beq
G \wedge G = \ast G \wedge F^i = 0 \,,
\label{consist}
\eeq
where $F^i$ is the $SU(2)$ field strength and $\ast G$ is the Hodge dual
of $G$ in eight dimensions \footnote{For a similar discussion in $SO(4)$
seven dimensional gauged supergravity, see \cite{gr}.}.

The D6-brane configurations we will be dealing with have some flux of $G$
along the directions of the worldvolume which are not wrapped. We will
first obtain them by looking at the supersymmetric transformations of the
fermionic fields which, after implementing the topological twisting and
imposing some projection conditions on the supersymmetric parameter, give
rise to some first-order differential equations for the metric and scalar
fields. We will obtain the same first-order equations by looking at the
effective lagrangian for our ansatz and verifying that its potential can
be derived from a superpotential, whose associated {\em domain wall}
equations are precisely those obtained from supersymmetry. Moreover, we
will be able to find the general solution to these equations and the
corresponding supergravity backgrounds in D=11 will be completely
determined. We end this section by considering the KK reduction of our
solution and several of its duals.

\subsection{First-order equations from SUSY}

Let us consider a D6--brane wrapping a special Lagrangian $3$--cycle of
the deformed conifold $T^*S^3$ from the point of view of eight-dimensional
gauged supergravity. We will switch a 4-form $G$ flux along three of the
unwrapped directions of the brane and the radial direction. The presence
of this flux introduces a distinction between one of the unwrapped
directions of the brane and the other three. Accordingly, the ansatz for
the metric will be:
\beq
ds^2_8\,=\,e^{2f}\,dx_{1,2}^2\,+\,e^{2\alpha}\,dy^2\,+\,
e^{2h}\,d\Omega_3^2\,+\,dr^2\,\,,
\label{uno}
\eeq
where $d\Omega_3^2$ is the metric of the unit $S^3$,  $f$, $\alpha$ and $h$
are functions of the radial coordinate $r$ to be determined and
$dx_{1,2}^2=-(dx^{0})^2+(dx^{1})^2+(dx^{2})^2$. The corresponding ansatz
for the 4-form $G$ in flat coordinates is:
\beq
G_{\underline{x^{0}x^{1}x^{2}r}}\,=\,\Lambda\,e^{-\alpha-3h-2\phi}\,\,,
\label{dos}
\eeq
with $\Lambda$ being a constant and $\phi$ the eight-dimensional dilaton.
This ansatz for $G$ ensures that its equation of motion is satisfied. Also,
the first equation in (\ref{consist}) is satisfied. Moreover, we will
parametrize the $S^3$ by means of the left invariant 1-forms $w^i$ on
the $SU(2)$ group manifold satisfying:
\beq
dw^i\,=\,{1\over 2}\,\epsilon_{ijk}\,w^j\,\wedge\,w^k\,\,.
\label{tres}
\eeq
In terms of three Euler angles $\theta, \phi$ and $\psi$, the $w^i$'s are:
\bear
w^1&=&\cos\phi \,d\theta\,+\,\sin\theta\sin\phi\, d\psi\,\,,\rc
w^2&=&\sin\phi\, d\theta\,-\,\sin\theta\cos\phi\, d\psi\,\,,\rc
w^3&=&d\phi\,+\,\cos\theta\, d\psi\,\,,
\label{cuatro}
\eear
while the metric of the unit 3-sphere is:
\beq
d\Omega_3^2\,=\,{1\over 4}\,\sum_{i=1}^3\,(\,w^i\,)^2\,\,.
\label{cinco}
\eeq
A bosonic configuration of fields is supersymmetric iff the supersymmetry
variation of the fermionic fields, evaluated on the configuration, vanishes. In
our case the fermionic fields are two pseudo Majorana spinors $\psi_{\lambda}$
and $\chi_i$ and their supersymmetry transformations are:
\bear
\delta\psi_{\lambda}&=&D_{\lambda}\,\epsilon\,+\,{1\over 24}\,e^{\phi}\,
F_{\mu\nu}^{i}\,\Gamma_i\,(\,\Gamma_{\lambda}^{\mu\nu}\,-\,
10\,\delta_{\lambda}^{\mu}\,\Gamma^{\nu}\,)\,\epsilon\,-\,{g\over 288}\,
e^{-\phi}\epsilon_{ijk}\,\Gamma^{ijk}\Gamma_{\lambda}\,T\epsilon\,-\rc\rc
&&-{1\over 96}\,e^{\phi}\,G_{\mu\nu\rho\sigma}\,(\,
\Gamma_{\lambda}^{\mu\nu\rho\sigma}\,
-\,4\delta_{\lambda}^{\mu}\,\Gamma^{\nu\rho\sigma}\,)\,\epsilon\,\,,\rc\rc
\delta\chi_i&=&{1\over 2}\,(P_{\mu ij}\,+\,{2\over 3}\,\delta_{ij}\,
\partial_{\mu}\phi\,)\,\Gamma^j\,\Gamma^{\mu}\,\epsilon\,-\,
{1\over 4}\,e^{\phi}\,F_{\mu\nu i}\,\Gamma^{\mu\nu}\,\epsilon\,-\,
{g\over 8}\,e^{-\phi}\,(\,T_{ij}\,-\,{1\over 2}\,\delta_{ij}\,T\,)\,
\epsilon^{jkl}\Gamma_{kl}\epsilon\,-\,\rc\rc
&&-{1\over 144}\,e^{\phi}\,G_{\mu\nu\rho\sigma}\,\Gamma_i\,
\Gamma^{\mu\nu\rho\sigma}\,\epsilon\,\,.
\label{seis}
\eear
In eqs.(\ref{seis}) the $\Gamma$'s are $32\times 32$ Dirac matrices and
$T^{ij}$ parametrizes the potential energy of the $SL(3,\RR)/SO(3)$
scalars of the theory, $L_\alpha^i$, with $T=\delta_{ij}T^{ij}$ (see
appendix A). The covariant derivative is:
\beq
D\,\epsilon\,=\,\big(\,\partial\,+\,{1\over 4}\,
\omega^{ab}\,\Gamma_{ab}\,+\,{1\over 4}\,
Q_{ij}\,\Gamma^{ij}\,\big)\,\epsilon\,\,,
\label{siete}
\eeq
where $\omega^{ab}$ are the components of the spin connection and
$Q_{ij}$ is an antisymmetric matrix constructed from the scalars and the
$SU(2)$ gauge potential (see appendix A).

Following ref.\cite{en}, we will adopt an ansatz for the gauge field which
corresponds to a complete identification of the spin connection with the
R-symmetry gauge potential of the theory, namely:
\beq
A^i\,=\,-{1\over 2g}\,  w^i\,\,.
\label{ocho}
\eeq
The corresponding  field strength is $F^i\,=\,-{1\over 8g}\,\epsilon^{ijk}
\,w^j\,\wedge\,w^k$. In this case it is possible to get rid of the coset
scalars, $L_\alpha^i$; $T_{ij}$ and $Q_{ij}$ simply reducing to
$T_{ij}\,=\,\delta_{ij}$ and $Q_{ij}\,=\,-g\epsilon_{ijk}\,A^k$. For
convenience, we will use the following representation of the Clifford algebra
\beq
\Gamma^{\underline a}\,=\,\gamma^{\underline a}\,\,\otimes\,\II\,\,,
\,\,\,\,\,\,\,\,\,\,\,\,\,\,\,\,\,\,\,\,\,\,
\Gamma^{\underline i}\,=\,\gamma_9\,\,\otimes\,\sigma^i\,\,,
\label{once}
\eeq
where $\gamma^{\underline a}$ are eight dimensional Dirac matrices,
$\sigma^i$ are Pauli matrices and  $\gamma_9\,=\,i \gamma^{\underline 0}\,
\gamma^{\underline 1}\,\cdots\,\gamma^{\underline 7}$ ($\gamma_9^2\,=\,1$).
The first-order equations we are looking for are obtained by requiring the
vanishing of the right-hand side of eqs.(\ref{seis}). This is achieved
after imposing some projection conditions on the supersymmetry parameter
$\epsilon$, which reduce the amount of supersymmetry to some fraction of
that of the vacuum. The standard projections corresponding to the
D6--branes wrapping the $S^3$ \cite{en}:
\beq
(\,\gamma_{{\underline r}}\,\otimes \II\,)\,\epsilon=
\,-i\,(\,\gamma_9\otimes \II)\,\epsilon\,,
\,\,\,\,\,\,\,\,\,\,\,\,\,\,\,\,\,\,\,\,\,\,\,\,
(\,\gamma_{\underline{ab}}\,\otimes \II\,)\,\epsilon=
\,-(\,\II\otimes\sigma^{ab}\,)\,\epsilon\,\,,
\label{doce}
\eeq
have to be supplemented by a new one due to the presence of the $G$ flux:
\beq
(\,\gamma_{\underline{x^0x^1x^2}}\,\otimes \II\,)\,\epsilon\,=
\,-\epsilon\,\,.
\label{trece}
\eeq
The number of supercharges unbroken by this configuration is then one
half of those corresponding to the case $\Lambda = 0$, \ie\ two.
It is now straightforward to get the following BPS equations:
\bear
f'&=&-{1\over 2g}\,e^{\phi-2h}\,+\,{g\over 8}\,e^{-\phi}\,+\,{\Lambda\over 2}\,
e^{-\phi-3h-\alpha}\,\,,\rc
\alpha'&=&-{1\over 2g}\,e^{\phi-2h}\,+\,{g\over 8}\,e^{-\phi}\,-\,{\Lambda\over
2}\, e^{-\phi-3h-\alpha}\,\,,\rc
h'&=&{3\over 2g}\,e^{\phi-2h}\,+\,{g\over 8}\,e^{-\phi}\,-\,{\Lambda\over 2}\,
e^{-\phi-3h-\alpha}\,\,,\rc
\phi'&=&-{3\over 2g}\,e^{\phi-2h}\,+\,{3g\over 8}\,e^{-\phi}\,-\,{\Lambda\over
2}\, e^{-\phi-3h-\alpha}\,\,.
\label{catorce}
\eear
Notice that, as it should be, eqs.(\ref{catorce}) reduce to the first-order
equations found in ref.\cite{en} when $\Lambda=0$.

\medskip
\subsection{First-order equations from a superpotential}

Before finding the general integral of the BPS equations (\ref{catorce}), let
us derive them again by means of an alternative method which consists in
finding a superpotential for the effective lagrangian constructed out of the
introduction of our ansatz into the Lagrangian of D=8 gauged supergravity.
Actually (see appendix B), the equations of motion of eight dimensional
supergravity for our ansatz can be derived from the effective Lagrangian:
\bear
{\cal L}_{eff}&=&e^{3f+\alpha+3h}\,\Big[\,(f'\,)^2\,+\,
(h'\,)^2\,-\,{1\over 3}\,(\phi'\,)^2\,+\,
3\,f'\,h'\,+\,f'\,\alpha'\,+\,\alpha'\,h'\,\rc
&&+\,e^{-2h}\,+\,{g^2\over 16}\,e^{-2\phi}\,-\,
{1\over g^2}\,e^{2\phi-4h}\,-\,{\Lambda^2\over 3}\,
e^{-2\alpha\,-\,6h\,-\,2\phi}\,\Big]\,\,.
\label{quince}
\eear
Let us now introduce a new radial variable $\hat r$, whose relation to our
original coordinate $r$ is given by:
\beq
{dr\over d\hat r}\,=\,e^{-{3\over 2}\,h\,-\,{1\over 2}\alpha}\,\,.
\label{dseis}
\eeq
The lagrangian in the new variable is $\hat {\cal L}_{eff}\,=\,e^{-{3\over
2}\,h\,-\,{1\over 2}\alpha}\,{\cal L}_{eff}\,$, where we have taken into
account the corresponding jacobian. If we define a new scalar field
$\varsigma \equiv f\,+\,{1\over 2}\,\alpha\,+{3\over 2}\,h$, and let
the dot denotes differentiation with respect to $\hat r$, the effective
lagrangian takes the form:
\beq
\hat {\cal L}_{eff}\,=\,e^{c_1\varsigma}\Big[\,c_2\,\dot\varsigma\,^2
\,-{1\over 2}\,G_{ab}\,\,\dot\varphi^a\,\dot\varphi^b\,-\,V(\varphi)\,
\,\Big]\,\,,
\label{dnueve}
\eeq
where $c_1=3$, $c_2=1$, and $\varphi^a$ denotes a vector whose components
are $\alpha$, $h$ and $\phi$. The non-vanishing elements of the metric
$G_{ab}$ are $G_{\alpha\,\alpha}\,=\,G_{\alpha\,h}\,=\,{1\over 2}$,
$G_{h\,h}\,=\,{5\over 2}$ and $G_{\phi\,\phi}\,=\,{2\over 3}$.
The potential $V(\varphi)$ appearing in $\hat {\cal L}_{eff}$ is:
\beq
V(\varphi)\,=\,
{1\over g^2}\,e^{2\phi\,-\,7h\,-\,\alpha}\,-\,{g^2\over 16}\,
e^{-2\phi\,-\,3h\,-\,\alpha}\,-\,e^{-\,5h\,-\,\alpha}\,+\,
{\Lambda^2\over 3}\,e^{-2\phi\,-\,9h\,-\,3\alpha}\,\,.
\label{vuno}
\eeq
For an effective lagrangian as in (\ref{dnueve}) being supersymmetric, the
corresponding potential (\ref{vuno}) must originate from a superpotential
$W$ as:
\beq
V\,=\,{1\over 2}\,\,G^{ab}\,\,{\partial W\over \partial \varphi^a}\,
{\partial W\over \partial \varphi^b}\,-\,
{c_1^2\over 4c_2}\,\,W^2\,\,,
\label{vdos}
\eeq
where $G^{ab}$ is the inverse metric. In our particular case $W$ must satisfy:
\beq
V\,=\,{5\over 4}\,\Bigg(\,{\partial W\over \partial \alpha}\,\Bigg)^2\,+\,
{1\over 4}\,\Bigg(\,{\partial W\over \partial h}\,\Bigg)^2\,+\,
{3\over 4}\,\Bigg(\,{\partial W\over \partial \phi}\,\Bigg)^2\,-\,
{1\over 2}\,{\partial W\over \partial \alpha}\,{\partial W\over \partial h}
\,-\,{9\over 4}\,W^2\,\,.
\label{vcuatro}
\eeq
For the value of $V$ given above (eq.(\ref{vuno})) one can check that
eq.(\ref{vcuatro}) is satisfied by:
\beq
W\,=\,-{1\over g}\,e^{\phi\,-\,{7\over 2}\,h\,-\,{\alpha\over 2}}\,-\,
{g\over 4}\,e^{-\phi\,-\,{3\over 2}\,h\,-\,{\alpha\over 2}}\,+\,
{\Lambda\over 3}\,e^{-\phi\,-\,{9\over 2}\,h\,-\,{3\over 2}\,\alpha}\,\,.
\label{vcinco}
\eeq
It is now easy to verify that the first-order {\em domain wall} equations
for this superpotential:
\bear
\dot\varsigma&=& -{c_1\over 2c_2}\,W\,\,,\rc\rc
\dot\varphi^a&=&G^{ab}\,{\partial W\over \partial \varphi^b}\,\,.
\label{vtres}
\eear
are exactly the same (when expressed in terms of the old variable $r$) as
those obtained from the supersymmetric variation of the fermionic fields
(eqs.(\ref{catorce})).

\medskip
\subsection{Integration of the first-order equations }

Let us now integrate eqs.(\ref{catorce}). In order to simplify the
expressions that follow, we shall take from now on the coupling constant
$g=1$. It is rather easy to find a particular solution in which $h-\phi$
is constant. First of all, we change variables from $r$ to $t$, where $t$
is such that:
\beq
{dr\over dt}\,=\,e^{\phi}\,\,.
\label{vseis}
\eeq
It follows from the last two equations in (\ref{catorce}) that, if
$h-\phi$ does not depend on $t$, we must have:
\beq
\phi(t)\,=\,h(t)\,-{1\over 2}\,\log(12)\,\,.
\label{vsiete}
\eeq
By using eq.(\ref{vsiete}) it is not difficult to prove that the values
$f$, $\alpha$ and $h$ are:
\bear
f(t)&=&{t\over 12}\,\,-\,{1\over 4}\,
\log\Big(\,c\,+\,{12\Lambda\over 5}\,e^{-{5\over 6}t}\,\Big)\,,\rc\rc
\alpha(t)&=&{t\over 12}\,+\,{1\over 4}\,
\log\Big(\,c\,+\,{12\Lambda\over 5}\,e^{-{5\over 6}t}\,\Big)\,\,,\rc\rc
h(t)&=&{t\over 4}\,+\,{1\over 4}\,
\log\Big(\,c\,+\,{12\Lambda\over 5}\,e^{-{5\over 6}t}\,\Big)\,\,,
\label{vocho}
\eear
where $c$ is an integration constant. Requiring that $f=\alpha$ for
$\Lambda=0$ we fix this constant to the value $c=1$. Notice that this is
equivalent to a change of  variable in the coordinate $y$.

Let us now consider the general solution of eqs.(\ref{catorce}) in which
$\phi-h$ is not necessarily constant. First, we define the function
$x\,\equiv 12 \,e^{2\phi-2h}\,$. It follows from the first-order
equations (\ref{catorce}) that the differential equation satisfied
by $x$ is:
\beq
{dx\over dt}\,=\,{1\over 2}\,x\,(1\,-\,x)\,\,.
\label{treinta}
\eeq
The solution of this equation is:
\beq
x\,=\,{1\over 1+b\,e^{-{t\over 2}}}\,\,,
\label{tuno}
\eeq
$b$ being an integration constant. Notice that taking $b=0$ we get the
previous particular solution. In general, we get the following relation
between $\phi$ and $h$:
\beq
\phi(t)\,=\,h(t)\,-\,{1\over 2}\,\log(12)\,-\,
{1\over 2}\,\log\,\Big(\,1+b\,e^{-{t\over 2}}\,
\Big)\,\,.
\label{tdos}
\eeq
In order to integrate completely the system (\ref{catorce}), let us define
a new function $z\,\equiv\,e^{3h\,+\,\alpha}\,$. It is not difficult to
check that $z$ satisfies the following differential equation:
\beq
{dz\over dt}\,=\,\Big[\,{1\over 2}\,+\,{x\over 3}\,\Big]\,z\,-\,2\Lambda\,\,.
\label{tcuatro}
\eeq
Since $x$ is a known function of $t$, we can integrate $z(t)$. To express
the result of this integration, let us define the variable $s$ as:
\beq
s\equiv \Big[\,{1\over b}\,e^{{t\over 2}}\,+\,1\,\Big]^{{1\over 3}}\,\,,
\label{tcinco}
\eeq
and let $J(s)$ be the following indefinite integral:
\beq
J(s)\equiv \,-\,9\,\int{ds\over (s^3\,-\,1\,)^2}\,\,.
\label{tseis}
\eeq
By elementary methods one can perform the integration on the right-hand
side of (\ref{tseis}) and obtain an explicit expression for $J(s)$:
\beq
J(s)\,=\,{3s\over s^3-1}\,+\,2\sqrt{3}\,{\rm arccot}\,
\Big[\,{1+2s\over \sqrt{3}}\,\Big]
\,-\,\log\,\Big(1\,+\,{3s \over (s-1)^2}\,\Big)\,\,.
\label{tsiete}
\eeq
In terms of $J(s)$, the function $z$ is given by:
\beq
z\,=\,e^{{5t\over 6}}\,\Big(\,1\,+\,b\,e^{-{t\over 2}}\,\Big)^{{2\over 3}}\,
\,\Big(\,1\,+\,\tilde\Lambda\,J(s)\,\Big)\,\,,
\label{tocho}
\eeq
with $\tilde\Lambda\,=\,{4\over 3}\,b^{-{5\over 3}}\,\Lambda\,$.
In (\ref{tocho}) we have fixed the integration constants
to reproduce the $b=0$ solution.  From these results it is easy to
obtain  the remaining  functions in (\ref{catorce}). They are:
\bear
f(t)&=&{t\over 12}\,-\,{1\over 12}\log \Big(\,1\,+\,b\,e^{-{t\over 2}}
\,\Big)\,-\,{1\over 4}\,
\log\,\Big(\,1\,+\,\tilde\Lambda\,J(s)\,\Big)\,\,,\rc\rc
\alpha(t)&=&{t\over 12}\,-\,{1\over 12}\log \Big(\,1\,+\,b\,e^{-{t\over 2}}
\,\Big)\,+\,{1\over 4}\,
\log\,\Big(\,1\,+\,\tilde\Lambda\,J(s)\,\Big)\,\,,\rc\rc
h(t)&=&{t\over 4}\,+\,{1\over 4}\log \Big(\,1\,+\,b\,e^{-{t\over 2}}
\,\Big)\,+\,{1\over 4}\,
\log\,\Big(\,1\,+\,\tilde\Lambda\,J(s)\,\Big)\,\,,
\label{cuarenta}
\eear
while $\phi(t)$ can be obtained from eq. (\ref{tdos}).
In the limit $b\approx 0$, after taking into account that
$J\approx {9\over 5}\,b^{{5\over 3}}\,e^{-{5t\over 6}}$, one easily
verifies that the solution (\ref{vocho}) is recovered.

\subsection{Uplifting to eleven dimensions}

Let us now analyze the eleven dimensional background corresponding to the D=8
BPS configurations found above. The uplifting formula for the metric is given
in appendix A.

\subsubsection{Smeared M2--branes on the tip of a $G_2$ cone}
We shall consider first the particular solution
(\ref{vsiete})--(\ref{vocho}). It is convenient to change again the radial
coordinate, from $t$ to a new coordinate $\rho$, whose relation is as
follows:
\beq
e^{{t\over 2}}\,=\,{1\over 18}\,\,\,\rho^3\,\,.
\label{cuno}
\eeq
Notice that, clearly, $\rho\ge 0$. By substituting the solution of
eqs.(\ref{vsiete})--(\ref{vocho}) in eq.(\ref{apacinco}), one gets the
following metric in D=11:
\beq
ds^2_{11}\,=\,
\big[\,H(\rho)\,\big]^{-{2\over 3}}\,dx_{1,2}^2\,+\,
\big[\,H(\rho)\,\big]^{{1\over 3}}\,\Big[\,
dy^2\,+\,d\rho^2\,+\,\rho^2\,ds^2_6\,\Big]\,\,,
\label{cdos}
\eeq
where $ds^2_6$ is the metric of a compact Einstein manifold, $Y_6$, with
the topology of $S^3\times S^3$,
\beq
ds^2_6\,=\,{1\over 12}\,\sum_{i=1}^3\,(\,w^i)^2\,+\,{1\over 9}\,
\sum_{i=1}^3\,\big(\,\tilde w^i\,-\,{1\over 2}\,w^i\,\big)^2\,\,,
\label{ccinco}
\eeq
whereas $H(\rho)$ is an harmonic function in the transverse seven
dimensional cone over $Y_6$ --whose metric, $d\rho^2\,+\,\rho^2\,
ds^2_6$, has $G_2$ holonomy--,
\beq
H(\rho)\,=\,1\,+\,{k\over \rho^5}\,\,.
\label{ctres}
\eeq
with $k$ being:
\beq
k\,=\,{1296\over 5}\,\sqrt{3}\,\,
{\Lambda\over (12)^{{1\over 6}}}\,\,.
\label{ccuatro}
\eeq
As for the 4-form $G$, we use the uplifting formula \footnote{The factor
of two is needed to pass from the Salam--Sezgin conventions of eleven
dimensional supergravity to the more standard ones. Notice that $F$ is
the corresponding 4-form in D=11 supergravity.}:
\beq
F_{\underline{x^0x^1x^2\rho}}\,=\,2\,e^{{4\phi\over 3}}\,
G_{\underline{x^0x^1x^2r}}\,\,.
\label{cseis}
\eeq
this leading, in curved indices, to the expression
\beq
F_{x^0x^1x^2\rho}\,=\,\epsilon_{x^0x^1x^2}\,\partial_{\rho}\,
\big[\,H(\rho)\,\big]^{-1}\,\,,
\label{csiete}
\eeq
where $\epsilon_{x^0x^1x^2}$ is the completely anti--symmetric Levi--Civita
tensor of the `external' (in the compactification language) Minkowski space.
It is clear from the result of the uplifting that our solution corresponds to
a smeared distribution of M2--branes in the tip of the singular cone over
$S^3 \times S^3$ with a $G_2$ holonomy metric found in \cite{bs,gpp}.
Notice that the power of $\rho$ in the harmonic function (\ref{ctres}) is
the one expected within this interpretation. Furthermore, notice that the
relation between the 4-form and the {\em warp} factor of the eleven
dimensional metric (\ref{cdos}) is characteristic of M Theory
compactifications in eight dimensional manifolds \cite{bb}.

The somehow unusual appearance of a smeared configuration in this approach
deserves some comments. We should first remind that, even in the case of
flat D--branes, it is well known that D2--branes have a low energy range,
$g^2_{YM} < U < g^2_{YM} N^{1 \over 5}$, in which string theory is strongly
coupled but the eleven dimensional curvature is small, and the appropriate
description is given in terms of the supergravity solution of smeared (in
the eleventh circle direction) M2--branes \cite{imsy}. In other words, the
D=11 configuration obtained by uplifting the D2--brane solution is not the
standard localized M2--brane. This result also holds in presence of
D6--branes. In fact, a system of D2--branes localized on flat D6--branes
\footnote{Localized intersections and overlappings of D--branes have
been studied in \cite{local}.} always has a low energy range described
by smeared M2--branes \cite{ps}. It is natural to expect that, if the
D6--branes are wrapping a supersymmetric cycle, the corresponding
description will be given in terms of smeared M2--branes transverse to
some special holonomy manifold. When we go further towards the IR, say
$U < g^2_{YM}$, we expect the smeared solution to be replaced (resolved)
by a periodic array of localized M2--branes along the eleventh circle.
Closer enough to the M2--branes, we should recover a conformal field
theory. There must be a more physical solution in D=11 supergravity
smoothly describing this transition in which, flowing towards the UV,
the solution smears before the eleventh circle radius becomes smaller
than the eleven dimensional Planck length \cite{imsy}.

\subsubsection{Smeared M2--branes on the resolved $G_2$ cone}

The case considered above is singular. The general solution (\ref{cuarenta})
obtained before resolves the conical singularity of the transverse $G_2$
manifold. Let us then uplift that solution (\ref{cuarenta}). First of all,
we introduce the parameter $a$, related to the integration constant $b$ of
eq.(\ref{tuno}) as follows, $b\,=\,{a^3\over 18}$. Moreover, let us
further change to a new variable $\rho$, which is now related to $t$
by means of the expression:
\beq
e^{{t\over 2}}\,=\,{1\over 18}\,\big(\,\rho^3\,-\,a^3\,\big)\,\,.
\label{cnueve}
\eeq
It follows immediately from (\ref{cnueve}) that the range of $\rho$ is
$\rho\ge a$. On the other hand, the variable $s$ introduced in
eq.(\ref{tcinco}) is, in terms of $\rho$ and $a$, simply given by
$s\,=\,{\rho\over a}\,$. After some elementary calculations, one can
verify that the functions $f$, $\alpha$, $h$ and $\phi$~ of
eqs.(\ref{tdos}) and (\ref{cuarenta}) can be written as:
\bear
e^{2f}&=&{\rho\over (18)^{{1\over 3}}}\,\,
\Big(\,1\,-\,{a^3\over \rho^3}\,\Big)^{{1\over 2}}\,\,
\Big[\,H(\rho)\,\Big]^{-{1\over 2}},
\,\,\,\,\,\,\,\,\,\,\,\,\,\,\,\,\,\,\,\,
e^{2\alpha}={\rho\over (18)^{{1\over 3}}}\,\,
\Big(\,1\,-\,{a^3\over \rho^3}\,\Big)^{{1\over 2}}\,\,
\Big[\,H(\rho)\,\Big]^{{1\over 2}},\rc\rc\rc
e^{2h}&=&{\rho^3\over 18}\,\,
\Big(\,1\,-\,{a^3\over \rho^3}\,\Big)^{{1\over 2}}\,\,
\Big[\,H(\rho)\,\Big]^{{1\over 2}},
\,\,\,\,\,\,\,\,\,\,\,\,\,\,\,\,\,\,\,\,
e^{2\phi}={\rho^3\over 216}\,\,
\Big(\,1\,-\,{a^3\over \rho^3}\,\Big)^{{3\over 2}}\,\,
\Big[\,H(\rho)\,\Big]^{{1\over 2}},
\label{ciuno}
\eear
where now the harmonic function $H(\rho)$ is given by:
\beq
H(\rho)\,=\,e^{2(\alpha\,-\,f)}\,=\,1\,+\,\tilde\Lambda \,
J\big(\,{\rho\over a}\,\big)\,\,.
\label{cidos}
\eeq
By using the explicit value of the function $J$ (eq.(\ref{tsiete})), one
can obtain the expression of $H(\rho)$, namely:
\beq
H(\rho)\,=\,1+\,k\,\Bigg[\,{5\over 3 a^3\rho^2}\,
{1\over 1\,-\,{a^3\over \rho^3}}+{10\over  3\sqrt{3}\,a^5}\,
{\rm arccot}\,\big[\,{2\rho +a\over a\sqrt{3}}\big]\,
-\,{5\over 9a^5}\,\log\big(\,1\,+\,{3 a\rho\over
(\rho-a)^2}\big)\,\Bigg]\,\,,
\label{citres}
\eeq
where the constant $k$ is the same as in eq. (\ref{ccuatro}). The
uplifted D=11 metric is now of the form:
\beq
ds^2_{11}\,=\,
\big[\,H(\rho)\,\big]^{-{2\over 3}}\,dx_{1,2}^2\,+\,
\big[\,H(\rho)\,\big]^{{1\over 3}}\,\Big[\,
dy^2\,+\,ds^2_7\,\Big]\,\,,
\label{cicuatro}
\eeq
where $ds^2_7$ is the metric of a regular manifold of $G_2$ holonomy found
in \cite{bs, gpp}, which is topologically $\RR^4\times S^3$, namely:
\beq
ds^2_7\,=\,{d\rho^2\over 1\,-\,{a^3\over \rho^3}}\,+\,
{\rho^2\over 12}\,\sum_{i=1}^3\,(\,w^i)^2\,+\,{\rho^2\over 9}\,
\big(\,1\,-\,{a^3\over \rho^3}\,\big)\,\sum_{i=1}^3\,
\big(\,\tilde w^i\,-\,{1\over 2}\,w^i\,\big)^2\,\,,
\label{cicinco}
\eeq
while the 4-form is still given by (\ref{csiete}) (with $H(\rho)$ now being the
function (\ref{citres})). This solution represents a smeared distribution of
M2--branes on the resolved manifold of $G_2$ holonomy $X_7$ whose singular limit
is the cone over $Y_6$ obtained above. It is an $\RR^4$ bundle over $S^3$.
We see again that the effect of the 4-form flux on the metric is
just the introduction of the corresponding warp factors. Actually,
the function $H(\rho)$ can also be determined by solving the Laplace
equation on the seven dimensional $G_2$ manifold \cite{cghp}. It is
also interesting to analyze the large and small distance behavior of
this harmonic function. When $\rho\rightarrow\infty$, $H(\rho)$
can be approximated as:
\beq
H(\rho)\approx 1\,+\,{k\over \rho^5}\,+\,{5a^3k\over 4\rho^8}
\,+\,\cdots\,\,,
\label{ciseis}
\eeq
\ie\ it has the same  leading asymptotic behaviour as the function
(\ref{ctres}). On the other hand, for $\rho\approx a$, $H(\rho)$
diverges as:
\beq
H(\rho)\approx {5k\over 9a^4}\,{1\over \rho-a}\,+\,
{10k\over 9a^5}\,\log{\rho-a\over a}\,+\,\cdots\,\,.
\label{cisiete}
\eeq
It is tempting to argue at this point that this supergravity smeared
solution might be the dual of some gauge theory at a given low energy
range. The resolution of the conical singularity must render the theory
non-conformal in the IR. In order to better understand our solutions,
it is important to go to ten dimensions. There are different reductions
to type IIA string theory: we can reduce on the smeared direction, or
we can embed the M--theory circle in the $\RR^4$ fiber or the $S^3$
base in $X_7$. We will study them in the following subsection.

\medskip
\subsection{Reduction to D=10 and T-duality}

Given an eleven dimensional metric with a Killing vector $v$, one can
generate a background of type IIA D=10 supergravity by means of a
Kaluza-Klein dimensional reduction along the direction of $v$. Actually,
if $\partial/\partial z$ is such a Killing vector, the reduction ansatz
for the metric is:
\beq
ds^2_{11}\,=\,e^{-{2\over 3}\,\phi_D}\,ds^2_{10}\,+\,
e^{{4\over 3}\,\phi_D}\,(\,dz\,+\,C^{(1)}\,)^2\,\,,
\label{ciocho}
\eeq
where $\phi_D$ is the ten-dimensional dilaton and $C^{(1)}$ is the RR
potential 1-form of the type IIA theory. In the case of our D=11 metric
(\ref{cicuatro}), we have several possibilities to choose $z$.

\subsubsection{D2--branes on the tip of a (resolved) $G_2$ cone}

The simplest election --and the most meaningful from the point of view of
gauge/string duality, as long as the smearing is removed-- is $z=y$, for
which the metric and dilaton of the IIA theory are:
\bear
ds^2_{10}&=&
\big[\,H(\rho)\,\big]^{-{1\over 2}}\,dx_{1,2}^2\,+\,
\big[\,H(\rho)\,\big]^{{1\over 2}}\,ds^2_7\,\,,\rc\rc
e^{\phi_D}&=&\big[\,H(\rho)\,\big]^{{1\over 4}}\,\,,
\label{cinueve}
\eear
while the 4-form field strength of D=11 becomes the RR 4-form $F^{(4)}$ of
the type IIA theory and $C^{(1)}$ vanishes. It is clear that this D=10
solution represents a D2 sitting at the tip of the $G_2$ holonomy
manifold $X_7$, whose principal orbits are topologically trivial $\tilde
S^3$ bundles over $S^3$. In the singular limit, when the base $S^3$ has
vanishing volume, we end with D2--branes at the tip of the $G_2$ cone
over the Einstein manifold $Y_6$. This configuration is reminiscent of
the Klebanov--Witten's $D3$--branes placed at the tip of the conifold
\cite{kw}. Indeed, it is a sort of lower supersymmetric version of it.
Notice, however, that the solution resulting from gauged supergravity is
the complete D2--brane solution and not its near horizon limit. This
might look strange since gauged supergravity usually gives directly the
near horizon metric. The reason is that the near horizon limit of the
D6--branes (that we would obtain through a different reduction, see
below), which are the {\em host} branes of D=8 gauged supergravity,
do not imply, in general, the near horizon limit of the D2--branes that
are intersecting them. We will come back to this point later. In summary,
in order to get the supergravity dual of the system of D2--branes on the
tip of the $G_2$ cone, we must consider the near horizon limit. We should
reintroduce $l_p$ units everywhere and take $\rho$, $a$ and $l_p$ to zero
such that
\beq
U \equiv {a\rho\over l_p^3} ~~~~~~~ \mbox{and}
~~~~~~~ L \equiv {a^2\over l_p^3}
\label{nearh}
\eeq
are kept fixed. The resulting expression for the harmonic function
(\ref{citres}), for large $U$, admits the following asymptotic expansion
\beq
H(U)\,=\,{5\, g_{YM}^3\, N\over 3 \, l_s^4\, L^3\, U^2} \sum_{n=1}^\infty
{3n\over 3n+2} \Big( {L\over U} \Big)^{3n} \,,
\label{harexp}
\eeq
where $g_{YM}^2 \approx L$ is the three dimensional coupling constant,
$a l_s^2 = l_p^3$, and $N$ is the number of $D2$--branes. The asymptotic
background gives the near horizon limit of $N$ $D2$--branes transverse
to the $G_2$ holonomy manifold:
\bear
ds^2_{10} & = & l_s^2 \left( {U^{5 \over 2}\over \sqrt{g_{YM}^2 N}}
\,dx_{1,2}^2\,+\, {\sqrt{g_{YM}^2 N} \over U^{{5\over 2}}} \,ds^2_7\,
\right) \,,\rc\rc
e^{\phi_D} & = & \left( {g_{YM}^{10} N \over U^5} \right)^{{1\over 4}}\,,
\label{cinuevenh}
\eear
and the 4-form field strength $F$ is still given by (\ref{csiete}).
It is analogous to the flat D2--brane \cite{imsy} except for the fact
that the transverse $\RR^7$ has been replaced by the $G_2$ cone over $S^3
\times S^3$. This is the valid description for intermediate high energies,
$g_{YM}^2 N > U > g_{YM}^2 N^{1\over 5}$, where the string coupling and
the curvature are small, and the radius of the eleventh circle vanishes.

In the UV we can trust the super Yang--Mills theory description. It is an
${\cal N}=1$ theory in $2+1$ dimensions. We can obtain its field content
following the arguments in \cite{kw}. In the case of a single $D2$--brane,
it is a $U(1) \times U(1)$ gauge theory with four complex scalars $Q_i$,
$\tilde Q_i$, $i=1,2$, and a vector multiplet whose gauge field can be
dualized to a compact scalar that would parametrize the position of the
$D2$--branes along the M--theory circle. The vacuum moduli space is given
by
\beq
|q_1|^2 + |q_2|^2 - |\tilde q_1|^2 - |\tilde q_2|^2 = L^2 ~,
\label{vmoduli}
\eeq
where $q_i$, $\tilde q_i$ are the scalar components of the superfields
$Q_i$, $\tilde Q_i$, which precisely provides an algebraic--geometric
description of the manifold $X_7$ \cite{amv}.

\subsubsection{D2--D6 system wrapping a special Lagrangian $S^3$}

The second possibility we shall explore is the reduction along some compact
direction of the $G_2$ manifold. Let us consider first the three-sphere $\tilde
S^3$, parametrized by the $su(2)$ left-invariant 1-forms $\tilde w^i$. Notice
that $\tilde S^3$ is external to the D6-brane worldvolume in the D=8 gauged
supergravity approach. We shall regard the $\tilde S^3$ sphere as a Hopf
bundle over a two-sphere, and we will reduce along the fiber of this bundle.
Let us denote by $\tilde \phi$, $\tilde \theta$ and $\tilde \psi$ the angles
which parametrize the $\tilde w^i$'s, as in eq.(\ref{cuatro}) after putting
tildes on both sides of the equation. We shall choose $z=\tilde \psi$ as the
coordinate along which the dimensional reduction will take place. Accordingly
\cite{clp}, let us  define the vector $\tilde \mu^i$ and the 1-forms
$\tilde e^i$ by means of the following decomposition of the $\tilde w^i$'s:
\beq
\tilde w^i\,=\,\tilde e^i\,+\,\tilde \mu^i\,d\tilde \psi\,\,.
\label{sesenta}
\eeq
The components of $\tilde \mu^i$ and $\tilde e^i$ are:
\bear
&&\tilde \mu^1\,=\,\sin\tilde\theta\sin\tilde\phi\,\,,
\,\,\,\,\,\,\,\,\,\,
\tilde \mu^2\,=\,-\sin\tilde\theta\cos\tilde\phi\,\,,
\,\,\,\,\,\,\,\,\,\,
\tilde \mu^3\,=\,\cos\tilde\theta\,\,,\rc\rc
&&\tilde e^1\,=\,\cos\tilde\phi\, d\tilde\theta\,\,,
\,\,\,\,\,\,\,\,\,\,\,\,\,\,\,\,\,
\tilde e^2\,=\,\sin\tilde\phi\, d\tilde\theta\,\,,
\,\,\,\,\,\,\,\,\,\,\,\,\,\,\,\,\,\,\,\,\,\,\,
\tilde e^3\,=\,d\tilde\phi\,\,.
\label{suno}
\eear
Notice that $\tilde\mu^i\tilde\mu^i\,=\,1$. One can also check the following
relation:
\beq
\tilde e^i\,=\,\epsilon_{ijk}\,\tilde \mu^j\,d\tilde \mu^k\,+\,
\cos\tilde\theta\,d\tilde\phi\,\tilde \mu^i\,\,,
\label{sdos}
\eeq
{}from which it follows that
$\tilde e^i\tilde\mu^i\,=\,\cos\tilde\theta\,d\tilde\phi$. Next, let us define
the one-forms $D\tilde \mu^i$ as:
\beq
D\tilde \mu^i\,\equiv\,d\tilde \mu^i\,-\,{1\over 2}\,\epsilon_{ijk}\,
w^j\,\tilde \mu^k\,\,.
\label{scuatro}
\eeq
It is important to point out  that the $D\tilde \mu^i$ one-forms are not
independent since $\tilde\mu^i\,D\tilde \mu^i\,=\,0$. Moreover, after some
calculation one verifies \cite{clp} that:
\beq
\,\sum_{i=1}^3\,\big(\,\tilde w^i\,-\,{1\over 2}\,w^i\,\big)^2\,=\,
\,\sum_{i=1}^3\,\big(\,D\tilde \mu^i\,\big)^2\,+\,\sigma^2\,\,,
\label{scinco}
\eeq
where $\sigma$ is given by:
\beq
\sigma\,=\,d\tilde \psi\,+\,\cos\tilde\theta\,d\tilde\phi\,-\,
{1\over 2}\,\tilde \mu^i\,w^i\,\,.
\label{sseis}
\eeq
Using eq. (\ref{scinco}) to rewrite the right-hand side of  (\ref{cicinco}),
one is able to put the metric (\ref{cicuatro}) in the form (\ref{ciocho}) with
$z=\tilde\psi$. Before giving the form of the resulting D=10 supergravity
background, let us write a more explicit expression for
$\big(\,D\tilde \mu\,\big)^2$,
\bear
\,\sum_{i=1}^3\,\big(\,D\tilde \mu^i\,\big)^2&=&\Bigg(\,d\tilde\theta\,-\,
\cos\tilde\phi\,{w^1\over 2}\,-\,\sin\tilde\phi\,{w^2\over 2}
\,\Bigg)^2\,\rc\rc
&&+\,\sin^2\tilde\theta\,\Bigg(\,d\tilde\phi\,+\,
\cot\tilde\theta\,\sin\tilde\phi\,{w^1\over 2}\,-\,
\cot\tilde\theta\,\cos\tilde\phi\,{w^2\over 2}\,-\,
{w^3\over 2}\Bigg)^2\,.
\label{ssiete}
\eear
If we  define $\gamma(\rho)$ as:
\beq
\gamma(\rho)\,\equiv\,{\rho^2\over 9}\,\,\big(\,1\,-\,{a^3\over
\rho^3}\,\big)\,\,,
\label{socho}
\eeq
then, the D=10 metric and dilaton obtained by reducing along $\tilde\psi$ are:
\bear
ds^2_{10}&=&\Bigg[\,{\gamma(\rho)\over H(\rho)}\,\Bigg]^{{1\over 2}}\,
\Big[\,dx^2_{1,2} + H(\rho)\,\big(\,dy^2\,+\,
{d\rho^2\over 1\,-\,{a^3\over \rho^3}}\,+\,{\rho^2\over 12}\,\sum_{i=1}^3\,
(\,w^i\,)^2\,+\,\gamma(\rho)\,\sum_{i=1}^3\,\big(\,D\tilde \mu^i\,\big)^2
\,\big)\,\Big]\,,\rc\rc
e^{\phi_D}&=&\Big[\,\gamma(\rho)\,\Big]^{{3\over 4}}\,
\Big[\,H(\rho)\,\Big]^{{1\over 4}}\,.
\label{snueve}
\eear
As the dilaton $\phi_D$ diverges at $\rho\rightarrow\infty$, it follows
that this solution has infinite string coupling constant. Moreover, the
RR potentials $C^{(1)}$ and $C^{(3)}$ of the type IIA theory are:
\bear
C^{(1)}&=&\cos\tilde\theta\,d\tilde\phi\,
-\,{1\over 2}\,\tilde\mu^i\,w^i\,\,,\rc\rc
C^{(3)}&=&-\Big[\,H(\rho)\,\Big]^{-1}\,\,
dx^0\wedge dx^1\wedge dx^2\,\,,
\label{setenta}
\eear
whose field strengths are:
\bear
F^{(2)}&=&-{1\over 2}\,\epsilon_{ijk}\,\tilde\mu^k\,
\Big[\,D\tilde\mu^i\wedge D\tilde\mu^j\,+\,{1\over 4}\,
w^i\wedge w^j\,\Big]\,\,,\rc\rc
F^{(4)}&=&\partial_{\rho}\,\Big[\,H(\rho)\,\Big]^{-1}\,\,
dx^0\wedge dx^1\wedge dx^2\wedge d\rho\,\,,
\label{stuno}
\eear
which clearly correspond to a (D2-D6)-brane system with the D2--brane
smeared in one of the directions of the D6--brane worldvolume (\ie\ along
the $y$ direction). Three of the directions of the D6--brane are
wrapping a supersymmetric 3-cycle in a complex deformed Calabi--Yau.
Yet, the smearing in D=10 makes this solution a bit awkward from the
point of view of the AdS/CFT correspondence. Instead, we can perform a
T--duality transformation along that direction.

\subsubsection{Curved D3--branes and deformed conifold}

Notice that $\partial/\partial y$ is still a Killing vector of the D=10
metric (\ref{snueve}). Therefore, we can perform a T-duality transformation
along the direction of the coordinate $y$ and, in this way, we get the
following solution of the type IIB theory:
\bear
ds^2_{10}&=&\Bigg[\,{\gamma(\rho)\over H(\rho)}\,\Bigg]^{{1\over 2}}\,
\Big[\,dx^2_{1,2} + {dy^2\over\gamma(\rho)} +
H(\rho)\,\left( {d\rho^2\over 1\,-\,{a^3\over \rho^3}}\,+\,{\rho^2\over 12}
\sum_{i=1}^3\,(w^i)^2 + \gamma(\rho) \sum_{i=1}^3\,\big(\,D\tilde
\mu^i\,\big)^2 \,\right)\,\Big]\,,\rc\rc
e^{\phi_D}&=&\Big[\,\gamma(\rho)\,\Big]^{{1\over 2}}\,,\rc\rc
F^{(3)}&=&{1\over 2}\,\epsilon_{ijk}\,\tilde\mu^k\,
\Big[\,D\tilde\mu^i\wedge D\tilde\mu^j\,+\,{1\over 4}\,
w^i\wedge w^j\,\Big]\wedge dy\,,\rc\rc
F^{(5)}&=&\partial_{\rho}\,\Big[\,H(\rho)\,\Big]^{-1}\,
dx^0\wedge dx^1\wedge dx^2\wedge  dy\wedge d\rho\,+\,
{\rm Hodge\,\, dual}\,\,.
\label{stdos}
\eear
The solution (\ref{stdos}) contains a D3-brane extended along $(x^1,x^2,y)$,
with the $y$-direction distinguished from the other two. For large $\rho$
the space transverse to the D3-brane is topologically a cone over $S^3\times
S^2$. Moreover, since $\gamma(\rho)\rightarrow 0$ as $\rho\rightarrow a$,
the $S^2$ part of the transverse space shrinks to zero near $\rho=a$ and,
thus, the transverse space has the same topology as the deformed conifold.

\subsubsection{Type IIA background with RR fluxes}

Another possible reduction to the type IIA theory is obtained by choosing
the M-theory circle as the Hopf fiber of the three sphere $S^3$ (the one
parametrized by the one-forms $w^i\,$). In order to proceed in this way,
let us first rewrite  the seven dimensional metric (\ref{cicinco}) as:
\beq
ds^2_7\,=\,{d\rho^2\over 1\,-\,{a^3\over \rho^3}}\,+\,
{\rho^2\over 12}\,\xi(\rho)\,\,\sum_{i=1}^3\,(\,\tilde w^i\,)^2\,
+\,\beta(\rho)\,\,\sum_{i=1}^3\,\big(\,
w^i\,-\,{\xi(\rho)\over 2}\,\tilde w^i\,\big)^2\,\,.
\label{sttres}
\eeq
with $\xi(\rho)$ and $\beta(\rho)$ being:
\beq
\xi(\rho)\,\equiv\,{1-{a^3\over \rho^3}\over 1-{a^3\over 4\rho^3}}\,\,,
\,\,\,\,\,\,\,\,\,\,\,\,\,\,\,\,\,\,
\beta(\rho)\,\equiv\,{\rho^2\over 9}\,
\big(\,1\,-\,{a^3\over 4 \rho^3}\,\big)\,.
\label{stcuatro}
\eeq
As in eq.(\ref{sesenta}), we decompose $w^i$ as
$w^i\,=\,e^i\,+\,\mu^i\,d\psi$. The components of $e^i$ and $\mu^i$ are similar
to the ones written in eq.(\ref{suno}). Moreover, if we define the 1-forms
$D\mu^i$ as:
\beq
D\mu^i\,\equiv\,d\mu^i\,-\,{\xi(\rho)\over 2}\,\epsilon_{ijk}\,
\tilde w^j\,\mu^k\,\,,
\label{stcinco}
\eeq
then, one can easily find expressions of the type of
eqs.(\ref{scinco})--(\ref{sseis}) and the D=10 solution is readily
obtained. For the metric, dilaton and RR 1-form potential one gets:
\bear
ds^2_{10}&=&\Bigg[\,{\beta(\rho)\over H(\rho)}\,\Bigg]^{{1\over 2}}\,\,
\Big[\,dx^2_{1,2}\,+\,H(\rho)\,\big(\,dy^2\,+\,
{d\rho^2\over 1\,-\,{a^3\over \rho^3}}\,+\,{\rho^2\over 12}\,\xi(\rho)\,
(\,\tilde w^i\,)^2\,+\,\beta(\rho)\,\big(\,D\mu^i\,\big)^2
\,\,\big)\,\Big]\,\,,\rc\rc
e^{\phi_D}&=&\Big[\,\beta(\rho)\,\Big]^{{3\over 4}}\,\,
\Big[\,H(\rho)\,\Big]^{{1\over 4}}\,\,,\rc\rc
C^{(1)}&=&\cos\theta\,d\phi\,
-\,{\xi(\rho)\over 2}\,\mu^i\,\tilde w^i\,\,,
\label{stseis}
\eear
while the RR potential $C^{(3)}$ is the same as in eq. (\ref{setenta}).

\subsubsection{Curved D3--branes and resolved conifold}

We can make a T-duality transformation to the background
(\ref{stseis}) in the direction of the coordinate $y$. The resulting metric
and dilaton are:
\bear
ds^2_{10}&=&\Bigg[\,{\beta(\rho)\over H(\rho)}\,\Bigg]^{{1\over 2}}\,\,
\Big[\,dx^2_{1,2}\,+\,{dy^2\over\beta(\rho)} \,+\,
H(\rho)\,\big(\,
{d\rho^2\over 1\,-\,{a^3\over \rho^3}}\,+\,{\rho^2\over 12}\,\xi(\rho)\,
(\,\tilde w^i\,)^2\,+\,\beta(\rho)\,\big(\,D \mu^i\,\big)^2
\,\,\big)\,\Big]\,\,,\rc\rc
e^{\phi_D}&=&\Big[\,\beta(\rho)\,\Big]^{{1\over 2}}\,\,,
\label{extrastseis}
\eear
which for large $\rho$ corresponds, again,  to a D3-brane with a transverse
space with the topology of a cone over $S^3\times S^2$.
However, since $\xi(\rho)$ vanishes at
$\rho=a$, in this case the $S^3$  part of the cone shrinks to zero as
$\rho\rightarrow a$ and, therefore, the transverse space has a structure
similar to the resolved conifold.

\setcounter{equation}{0}
\section{D6-branes wrapped on a 2-sphere with 4-form}
\medskip

\subsection{First-order equations from SUSY}
In this section we will analyze the situation in which the
D6--branes are wrapped on a holomorphic two sphere and a four-form flux is
turned on along some of the unwrapped directions of its worldvolume. As
argued in ref.\cite{en}, one must excite in this case a real scalar field
--which parametrizes the Coulomb branch of the theory--, out of the coset
scalars $L_{\alpha}^{i}$ (see appendix A). We then adopt the following
ansatz for these scalars \cite{en}:
\beq
L_{\alpha}^{i}\,=\,
{\rm diag}\,(e^{\lambda}\,, \,e^{\lambda}\,,\, e^{-2\lambda})\,\,.
\label{stsiete}
\eeq
On the other hand, when the D6-brane wraps an $S^2$, two of the
unwrapped directions of its worldvolume are distinguished from the others by
the flux of the four-form. Thus, the natural ansatz for the metric in this
case is:
\beq
ds^2_8\,=\,e^{2f}\,dx_{1,2}^2\,+\,e^{2\alpha}\,dy_2^2\,+\,
e^{2h}\,d\Omega_2^2\,+\,dr^2\,\,,
\label{stocho}
\eeq
where $d\Omega_2^2$ is the metric of the unit $S^2$ and
$dy_2^2=(dy^1)^2+(dy^2)^2$. For the metric (\ref{stocho}), the equation of
motion of the four-form is satisfied if one adopts the following ansatz
for $G$:
\beq
G_{\underline{x^{0}x^{1}x^{2}r}}\,=\,\Lambda\,e^{-2\alpha-2h-2\phi}\,\,,
\label{stnueve}
\eeq
where, as before, $\Lambda$ is a constant. The BPS configurations for our
ansatz can be obtained by requiring the vanishing of the supersymmetry
variations of the fermionic fields. To find these configurations we must
determine first the spinor projections and the gauge field which implement
the appropriate topological twisting. In order to specify them, let us
represent the $S^2$ line element in terms of two angles $\theta^1$ and
$\phi^1$ as $d\Omega_2^2\,=\,(\,d\theta^1\,)^2+(\,\sin\theta^1\,)^2\,
(\,d\phi^1\,)^2$. The gauge field potential $A^i$ which we will consider
has only components along the direction $i=3$, its field strength being
given by the volume form of $S^2$ \cite{en},
\beq
A^3\,=\,{1\over g}\,\cos\theta^1 d\phi^1\,\,,
\label{ochenta}
\eeq
while the corresponding spinor projections are the ones in \cite{en} plus
an extra projection related to the presence of a $G$ flux:
\bear
&&(\,\gamma_{\underline{\theta^1\phi^1}}\otimes \II\,)\epsilon\,=
\,-(\,\II\otimes\sigma^1\,\sigma^2)\epsilon\,\,,
\,\,\,\,\,\,\,\,\,\,\,\,\,\,\,\,\,\,\,\,\,\,\,\,
(\,\gamma_{\underline{r}}\otimes \II\,)\,\epsilon=
\,-i(\,\gamma_9\otimes\II\,)\epsilon\,,\rc\rc
&&(\,\gamma_{\underline{x^0x^1x^2}}\,\otimes\II\,)\epsilon\,=
\,-\epsilon\,\,.
\label{ochentauno}
\eear
The number of unbroken supercharges is then four. 
It is now straightforward to find the first-order equations which follow from
the conditions $\delta\psi_{\lambda}=\delta\chi_i=0$. One gets:
\bear
f'&=&-{1\over 6g}\,e^{\phi-2h-2\lambda}\,
+\,{g\over 24}\,e^{-\phi}\,(\,2e^{2\lambda}\,+\,e^{-4\lambda}\,)
\,+\,{\Lambda\over 2}\, e^{-\phi-2h-2\alpha}\,\,,\rc
\alpha'&=&-{1\over 6g}\,e^{\phi-2h-2\lambda}\,
+\,{g\over 24}\,e^{-\phi}\,(\,2e^{2\lambda}\,+\,e^{-4\lambda}\,)
\,-\,{\Lambda\over 2}\, e^{-\phi-2h-2\alpha}\,\,,\rc
h'&=&{5\over 6g}\,e^{\phi-2h-2\lambda}\,
+\,{g\over 24}\,e^{-\phi}\,(\,2e^{2\lambda}\,+\,e^{-4\lambda}\,)
\,-\,{\Lambda\over 2}\, e^{-\phi-2h-2\alpha}\,\,,\rc
\phi'&=&-{1\over 2g}\,e^{\phi-2h-2\lambda}\,
+\,{g\over 8}\,e^{-\phi}\,(\,2e^{2\lambda}\,+\,e^{-4\lambda}\,)
\,-\,{\Lambda\over 2}\, e^{-\phi-2h-2\alpha}\,\,,\rc
\lambda'&=&{1\over 3g}\,e^{\phi-2h-2\lambda}\,
-\,{g\over 6}\,e^{-\phi}\,(\,e^{2\lambda}\,-\,e^{-4\lambda}\,)\,\,.
\label{ouno}
\eear
As a check, it is interesting to verify in eq. (\ref{ouno}) that, when
$\Lambda=0$,  $f'=\alpha'=\phi'/3$, and the resulting equations coincide
with those of ref.\cite{en}.

\medskip
\subsection{First-order equations from a superpotential}
It is also possible in this case to give another derivation of the first-order
system (\ref{ouno}) by analyzing the effective lagrangian for our ansatz. Let
us briefly present it here for completeness. After
some calculations one can verify that this effective lagrangian is:
\bear
{\cal L}_{eff}&=&e^{3f+2\alpha+2h}\,\Big[\,{3\over 2}\,\big(f\,'\,)^2\,+\,
{1\over 2}\,\big(\alpha\,'\,)^2\,+\,{1\over 2}\,\big(h\,'\,)^2\,-\,
{3\over 2}\,\big(\lambda\,'\,)^2\,-\,{1\over 2}\,\big(\phi\,'\,)^2\,\rc\rc
&&+\,3f\,'\,\alpha\,'\,+\,3f\,'\,h\,'\,+\,2h\,'\alpha\,'\,+\,
{1\over 2}\,e^{-2h}\,
+\,{g^2\over 16}\,e^{-2\phi}\,\big(\,2e^{-2\lambda}\,-
\,{1\over 2}\,e^{-8\lambda}\,\big)\,\rc\rc
&&-\,{1\over 2g^2}\,e^{2\phi-4h-4\lambda}\,-\,
{\Lambda ^2\over 2}\,e^{-4\alpha-2\phi-4h}\,\,\Big]\,\,.
\label{odos}
\eear
As in section 2.2, let us now define a new variable $\hat r$ as:
\beq
{dr\over d\hat r}\,=\,e^{-h\,-\,\alpha}\,\,.
\label{otres}
\eeq
After taking into account the jacobian for the change of variable
(\ref{otres}), one concludes that the effective lagrangian in the new
variable is
$\hat {\cal L}_{eff}\,=\,e^{-h\,-\,\alpha}\,{\cal L}_{eff}$. Moreover, if the
dot denotes differentiation with respect to $\hat r$, it is easy to check that
$\hat {\cal L}_{eff}$ can be put in the form (\ref{dnueve}), with
$\varsigma \,\equiv\, f\,+\,h\,+\,\alpha$. In eq.(\ref{dnueve}), the
constants $c_1$ and $c_2$ become $c_1=3$, $c_2=3/2$, and now $\varphi^a$
has  four components, namely, $\varphi^a\,=\,(\alpha, h, \phi, \lambda)$.
The non-vanishing elements of the metric $G_{ab}$ are $G_{\alpha\,\alpha}
\,=\,G_{h\,h}\,=\,2$, $G_{\alpha\,h}\,=\,G_{\phi\,\phi}\,=\,1$ and
$G_{\lambda\,\lambda}\,=\,3$, and the potential $V$ is given by:
\beq
V={1\over 2g^2}\,e^{2\phi-6h-2\alpha-4\lambda}\,+\,
{g^2\over 32}\,e^{-2\phi-2h-2\alpha}\,
\big(\,e^{-8\lambda}\,-\,4e^{-2\lambda}\,\big)\,-\,\rc\rc
{1\over 2}\,e^{-4h-2\alpha}\,+\,
{\Lambda ^2\over 2}\,e^{-6\alpha-2\phi-6h}\,\,.
\label{oseis}
\eeq
The corresponding superpotential $W$ must satisfy eq. (\ref{vdos}), which in
this case becomes:
\beq
V = {1\over 3}\,\Bigg(\,{\partial W\over \partial \alpha}\,\Bigg)^2\,+\,
{1\over 3}\,\Bigg(\,{\partial W\over \partial h}\,\Bigg)^2\,+\,
{1\over 2}\,\Bigg(\,{\partial W\over \partial \phi}\,\Bigg)^2\,+\,
{1\over 6}\,\Bigg(\,{\partial W\over \partial \lambda}\,\Bigg)^2\,-\,\rc\rc
{1\over 3}\,{\partial W\over \partial h}\,
{\partial W\over \partial \alpha}\,-\,{3\over 2}\,\,W^2\,\,.
\label{osiete}
\eeq
After some elementary calculation, one can prove that $W$ can be taken as:
\beq
W=-{1\over 2g}\,\,e^{\phi\,-\,3h\,-\,\alpha\,-\,2\lambda}\,-\,
{g\over 8}\,\,e^{-\phi\,-\,h\,-\,\alpha}\,\,
\big(\,e^{-4\lambda}\,+\,2e^{2\lambda}\,\big)\,
+\,{\Lambda\over 2}\,\,e^{-3\alpha-3h-\phi}\,\,.
\label{oocho}
\eeq
The first-order equations for this superpotential can be obtained by
substituting (\ref{oocho}) on the right-hand side of eq.(\ref{vtres}). It is
not difficult to check that, in terms of the original variable $r$, one gets
exactly the first-order system (\ref{ouno}).

\medskip
\subsection{Integration of the first-order equations }

We now undertake the task of integrating the system (\ref{ouno}). As in section
2.3, we will take $g=1$ from now on and we shall begin by a simpler particular
case, in which some combinations of the unknown functions are constant. By
inspecting (\ref{ouno}) one easily realizes that $\lambda$ can be kept constant
if $\phi-h$ is also constant. Actually, in this case one must have:
\beq
\lambda\,=\,{1\over 6}\,\,\log\,\Big(\,{3\over 2}\,\Big)\,\,,
\,\,\,\,\,\,\,\,\,\,\,\,\,\,\,\,\,\,\,\,\,\,\,\,
\phi\,=\,h\,-\,{1\over 6}\,\,\log\,(96)\,\,,
\label{onueve}
\eeq
as in the singular solution found in \cite{en} when $\Lambda = 0$.
In order to integrate completely the system in this particular case,  let us
change variables from $r$ to $t$, where $t$ is determined by the condition:
\beq
{dr\over dt}\,=\,e^{\phi\,+\,4\lambda}\,\,.
\label{noventa}
\eeq
Then, defining $\hat\Lambda\,=\,2\,\Big(\,{3\over 2}\,\Big)^{{2\over
3}}\,\Lambda$, we find the following solution:
\bear
f(t)\,&=&\,{1\over 8}\,t\,-\,{1\over 4}
\log\Big(\,1\,+\,\hat\Lambda\, e^{-t}\,)\,\,,\rc
\alpha(t)\,&=&\,{1\over 8}\,t\,+\,{1\over 4}
\log\Big(\,1\,+\,\hat\Lambda\, e^{-t}\,)\,\,,\rc
h(t)\,&=&\,{3\over 8}\,t\,+\,{1\over 4}
\log\Big(\,1\,+\,\hat\Lambda\, e^{-t}\,)\,\,,
\label{nuno}
\eear
where we have fixed the integration constants by imposing $f=\alpha$ for
$\Lambda=0$. Notice that $\phi$, in this solution, can be obtained from
(\ref{onueve}) and (\ref{nuno}).

Let us now find a general solution of (\ref{ouno}). First of all, we define the
function  $x\equiv 4e^{2\phi-2h+2\lambda}$. It can be easily verified that $x$
satisfies the differential equation (\ref{treinta}), where now $t$ is the
variable defined in (\ref{noventa}). We write the integral
of eq. (\ref{treinta})  as:
\beq
x\,=\,{1\over 1\,+\,ce^{-{t\over 2}}}\,\,,
\label{ndos}
\eeq
with $c$ being an integration constant. It follows from the first-order
system (\ref{ouno}) that $\lambda$ satisfies the equation:
\beq
{d\lambda\over dt}\,=\,{1\over 6}\,\big(\,1\,-\,e^{6\lambda}\,)\,
+\,{x\over 12}\,\,.
\label{ntres}
\eeq
By using the explicit dependence of $x$ on $t$, displayed in eq. (\ref{ndos}),
the integral of eq. (\ref{ntres}) is easy to find. In order to express this
integral in a convenient way, let us parametrize $\lambda$ as:
\beq
\lambda\,=\,{1\over 6}\,\Big[\,\log\big(\,{3\over 2}\,\big)
\,-\,\log \kappa\,\Big]\,\,.
\label{ncuatro}
\eeq
Notice that  $\kappa=1$ corresponds to the solution (\ref{onueve}), in which
$\lambda$ is not running and $x=1$. In general, the function $\kappa(t)$ is
given by:
\beq
\kappa(t)\,=\,{e^{{3\over 2}t}\,+\,{3\over 2}\,c\,e^{t}\,+\,d\over
e^{{3\over 2}t}\,+\,\,c\,e^{t}}\,\,,
\label{ncinco}
\eeq
where $d$ is a new integration constant. Next, let us define the function
$z$ as $z\equiv e^{2(\alpha+h)}$. After simple manipulations of the system
(\ref{ouno}), one reaches the conclusion that $z$ satisfies the equation:
\beq
{dz\over dt}\,=\,\Big[\,{x\over 3}\,+\,{1\over 6}\,
(\,2e^{6\lambda}\,+\,1\,)\,\Big]\,z\,-\,2\Lambda\,e^{4\lambda}\,\,,
\label{nseis}
\eeq
which can be solved by using the explicit dependence of $x$ and $\lambda$ on
$t$. To express the solutions of this equation, let us define the integral:
\beq
I(t)\equiv -\,\int\,
{dt\over e^t\,\big(\,1\,+\,{3\over 2}\,c\,e^{-{t\over 2}}\,
+\,d\,e^{-{3t\over 2}}\,\big)}\,\,.
\label{nsiete}
\eeq
Then, one has:
\beq
z\,=\,e^t\,
(\,1\,+\,c\,e^{-{t\over 2}}\,)\,\big[\,\kappa(t)\,\big]^{{1\over 3}}\,
(\,1\,+\,\hat\Lambda I(t)\,)\,\,,
\label{nocho}
\eeq
with $\hat\Lambda$ the same as in eq. (\ref{nuno}). Notice that in
(\ref{nocho}) we have fixed the integration constant to have the same
value of $z$ as in the solution  (\ref{nuno}) when $c=d=0$. Once $x$, $\lambda$
and $z$ are known, the functions $f$, $\alpha$, $h$ and $\phi$ can be obtained
by direct integration of the equations:
\bear
{df\over dt}&=&-{x\over 24}\,+\,{1\over 24}\,(\,2e^{6\lambda}\,+\,1\,)\,
+\,{\Lambda\over 2}\,{e^{4\lambda}\over z}\,\,,\rc\rc
{d\alpha\over dt}&=&-{x\over 24}\,+\,{1\over 24}\,(\,2e^{6\lambda}\,+\,1\,)\,
-\,{\Lambda\over 2}\,{e^{4\lambda}\over z}\,\,,\rc\rc
{dh\over dt}&=&{5x\over 24}\,+\,{1\over 24}\,(\,2e^{6\lambda}\,+\,1\,)\,
-\,{\Lambda\over 2}\,{e^{4\lambda}\over z}\,\,,\rc\rc
{d\phi\over dt}&=&-{x\over 8}\,+\,{1\over 8}\,(\,2e^{6\lambda}\,+\,1\,)\,
-\,{\Lambda\over 2}\,{e^{4\lambda}\over z}\,\,.
\label{nnueve}
\eear
The result is given as follows:
\bear
f(t)&=&{t\over 8}\,+\,{1\over 12}\,\log \kappa(t)\,-\,
{1\over 4}\,\log\big(\,1\,+\,\hat\Lambda I(t)\,\big)\,\,,\rc\rc
\alpha(t)&=&{t\over 8}\,+\,{1\over 12}\,\log \kappa(t)\,+\,
{1\over 4}\,\log\big(\,1\,+\,\hat\Lambda I(t)\,\big)\,\,,\rc\rc
h(t)&=&{3t\over 8}\,+\,{1\over 2}\,\log\big(\,1\,+\,ce^{-{t\over 2}}\,\big)
\,+\,{1\over 12}\,\log \kappa(t)\,+\,
{1\over 4}\,\log\big(\,1\,+\,\hat\Lambda I(t)\,\big)\,\,,\rc\rc
\phi(t)&=&{3t\over 8}\,+\,{1\over 4}\,\log \kappa(t)\,+\,
{1\over 4}\,\log\big(\,1\,+\,\hat\Lambda I(t)\,\big)
\,-\,{1\over 6}\log(96)\,\,,
\label{cien}
\eear
where, again, we have fixed the integration constants in order to reproduce the
solution (\ref{nuno}) when $c=d=0$.

\medskip
\subsection{Uplifting to eleven dimensions}

\subsubsection{Smeared M2--branes at the tip of the conifold}

Let us consider first the uplifting to eleven dimensions of the
particular solution (\ref{nuno}). Introducing a new radial coordinate
$\rho$ as:
\beq
e^{{t\over 2}}\,=\,{1\over 6(96)^{{1\over 9}}}\,\,\,\rho^2\,\,,
\label{ctuno}
\eeq
we get that  the corresponding eleven dimensional metric takes the form:
\beq
ds^2_{11}\,=\,
\big[\,H(\rho)\,\big]^{-{2\over 3}}\,dx_{1,2}^2\,+\,
\big[\,H(\rho)\,\big]^{{1\over 3}}\,\Big[\,
dy_2^2\,+\,d\rho^2\,+\,\rho^2\,ds^2_5\,\Big]\,\,,
\label{ctdos}
\eeq
where now the harmonic function $H(\rho)$ is:
\beq
H(\rho)\,=\,1\,+\,{k\over \rho^4}\,\,,
\label{cttres}
\eeq
and the constant $k$ is related to $\Lambda$ by means of the expression:
\beq
k\,=\,432\,\,
{\Lambda\over (96)^{{1\over 9}}}\,\,\,.
\label{ctcuatro}
\eeq
The metric $ds^2_5$ appearing in eq.(\ref{ctdos}) is the one corresponding to
the Einstein $T^{1,1}$ space, namely:
\beq
ds^2_5\,=\,{1\over 9}\,(\,d\psi\,+\sum_{a=1,2}\,\cos\theta_a\,d\phi_a\,)^2\,+\,
{1\over 6}\,\sum_{a=1,2}\,(\,d\theta_a^2\,+\,\sin^2\theta_a\,d\phi_a^2\,)\,\,.
\label{ctcinco}
\eeq
Recall that $d\rho^2\,+\,\rho^2\,ds^2_5$ is the metric of the singular
conifold with base $T^{1,1}$ \cite{cdlo}. The 4-form $F$ can be obtained
by plugging the solution (\ref{nuno}) into the uplifting formula
(\ref{cseis}). The result is just (\ref{csiete}), where now the harmonic
function $H(\rho)$ is the one given in (\ref{cttres}). It follows from
these results that this solution can be
interpreted as the geometry created by a smeared distribution of M2-branes
located at the tip of the singular conifold. Notice that we are now smearing
the M2- brane along two coordinates, which agrees with the power of $\rho$ in
eq. (\ref{cttres})

\subsubsection{Smeared M2--branes and generalized resolved conifold}

Let us now uplift the general solution (\ref{cien}). First of all we change
variables from $t$ to $\rho$ as in eq.(\ref{ctuno}). Then, let us represent
the constants $c$ and $d$ in terms of two parameters $a$ and $b$ as follows:
\beq
c\,=\,{1\over (96)^{{1\over 9}}}\,\,a^2\,\,,
\,\,\,\,\,\,\,\,\,\,\,\,\,\,\,\,\,\,\,\,\,\,
d\,=\,-{1\over 6^3(96)^{{1\over 3}}}\,\,b^6\,\,.
\label{ctseis}
\eeq
(we are assuming that $d\le 0$). With these definitions the function
$\kappa$ becomes:
\beq
\kappa(\rho)\,=\,{\rho^6\,+\,9a^2\,\rho^4\,-\,b^6\over
\rho^6\,+\,6a^2\rho^4}\,\,,
\label{ctsiete}
\eeq
and, if we define now:
\beq
H(\rho)\,=\,1\,+\,\hat\Lambda\,\,I(t(\rho))\,\,,
\label{ctocho}
\eeq
where $I(t(\rho))$ is the integral defined in eq.(\ref{nsiete}), then the
D=11 metric can be written as:
\beq
ds^2_{11}\,=\,
\big[\,H(\rho)\,\big]^{-{2\over 3}}\,dx_{1,2}^2\,+\,
\big[\,H(\rho)\,\big]^{{1\over 3}}\,\Big[\,
dy_2^2\,+\,ds^2_6\,\Big]\,\,,
\label{ctnueve}
\eeq
where the six dimensional metric $ds^2_6$ is given by:
\bear
ds^2_6\,&=&\,\big[\,\kappa(\rho)\,\big]^{-1}\,d\rho^2\,+\,
{\rho^2\over 9}\,\,\kappa(\rho)\,
(\,d\psi\,+\sum_{a=1,2}\,\cos\theta_a\,d\phi_a\,)^2\,+\,\rc\rc
&&+\,{1\over 6}\,(\,\rho^2\,+\,6a^2\,)\,
(\,d\theta_1^2\,+\,\sin^2\theta_1\,d\phi_1^2\,)\,+\,
{1\over 6}\,\rho^2\, (\,d\theta_2^2\,+\,\sin^2\theta_2\,d\phi_2^2\,)\,\,.
\label{ctdiez}
\eear
Moreover, by performing the change of variables (\ref{ctuno}) in
(\ref{nsiete}), we obtain the following integral representation of the
harmonic function $H(\rho)$:
\beq
H(\rho)\,=\,1\,+\,4k\,\int_{\rho}^{\infty}\,
{\tau d\tau\over \tau^6\,+\,9a^2\tau^4\,-\,b^6}\,\,,
\label{ctonce}
\eeq
with $k$ given by eq. (\ref{ctcuatro}). On the other hand, the 4-from $F$
for this solution can be put in the form (\ref{csiete}) with $H(\rho)$
given by eq.(\ref{ctonce}).

The six dimensional metric (\ref{ctdiez}) is the one corresponding to the small
resolution of the generalized conifold \cite{cdlo,pt,pzt1, pzt2}.
The parameter $b$ was introduced in refs.\cite{pt, pzt2}, where it
was pointed out that for $a=0$ and $b>0$ the metric (\ref{ctdiez}) can be made
non-singular if one takes $\rho\ge b$ and makes a $\ZZ_2$ identification of the
fiber coordinate $\psi$. In the context of D=8 gauged supergravity, the $b=0$
metric was obtained in ref.\cite{en}. Our result generalizes the one in
\cite{en}, even in the absence of 4-form flux, and shows that eight
dimensional gauged supergravity can easily incorporate the two-parameter
metric of refs.\cite{pt, pzt2}. Moreover, as we
have switched on the 4-form $F$ in our solution, the corresponding metric
contains the appropriate powers of the harmonic function $H(\rho)$, whose
integral representation is given in eq.(\ref{ctonce}). It is immediate to
conclude from this representation that $H(\rho)$ behaves for
$\rho\rightarrow\infty$ exactly as the right-hand side of eq.(\ref{cttres}).
In order to find out the behaviour of $H$ at small $\rho$, let us perform
explicitly the integral  (\ref{ctonce}) in some particular cases. First of all,
we consider the case $b=0$, for which $H(\rho)$ is given by:
\beq
H(\rho)\,=\,1\,+\,{2k\over 9a^2}\,{1\over \rho^2}\,-\,
{2k\over 81a^4}\,\log\big(\,1\,+\,{9a^2\over \rho^2}\,\big)\,\,,
\,\,\,\,\,\,\,\,\,\,\,\,\,\,\,
(b=0)\,\,.
\label{ctdoce}
\eeq
This  expression for $H(\rho)$ coincides exactly with the one found in
\cite{pzt1} for the case of a D3-brane at the tip of the small resolution
of the conifold, which can be obtained from our solution by dimensional
reduction and T-duality (see below). For $\rho\approx 0$ the harmonic
function behaves as:
\beq
H(\rho)\,\approx\,{2k\over 9a^2}\,{1\over \rho^2}\,\,,
\,\,\,\,\,\,\,\,\,\,\,\,\,\,\, (b=0)\,\,.
\label{cttrece}
\eeq
When $a=0$ the integral (\ref{ctonce}) can also explicitly performed , with
the result:
\beq
H(\rho)\,=\,1\,-\,{2k\over b^4}\,\,
\Big[\,{1\over 6}\,\log{(\rho^2-b^2)^3\over \rho^6-b^6}\,+\,{1\over \sqrt{3}}\,
{\rm arccot}{2\rho^2+b^2\over \sqrt{3}\,b^2}\,\Big]\,\,,
\,\,\,\,\,\,\,\,\,\,\,\,\,\,\, (a=0)\,\,,
\label{ctcatorce}
\eeq
and, again, this result coincides with that of ref.\cite{pzt2}. For
$\rho\approx b$ the function in (\ref{ctcatorce}) has a logarithmic behaviour
of the form:
\beq
H(\rho)\,\approx\,-{2k\over 3b^4}\,\log{\rho-b\over b}\,\,,
\,\,\,\,\,\,\,\,\,\,\,\,\,\,\, (a=0)\,\,.
\label{ctquince}
\eeq
For general values of $a$ and $b$ the
integral (\ref{ctonce}) can be  performed by factorizing the polynomial in
the denominator. The result depends on the sign of the ``discriminant"
$\Delta\,=\,b^6\,-\,108\,a^6$. The analysis of the different cases has
been carried out in ref. \cite{pzt2} and will not be repeated here.

\medskip
\subsection{Reduction to D=10 and T-duality}

As in section 2.5 we can dimensionally reduce and T-dualize the metric
(\ref{ctnueve}) along different directions.

\subsubsection{D3--branes at the tip of the generalized resolved conifold}

Let us consider first a reduction along a direction orthogonal to the six
dimensional metric (\ref{ctdiez}). Notice that $\partial/\partial y^1$ and
$\partial/\partial y^2$ are Killing vectors of (\ref{ctnueve}). Let us reduce
along $y^2$ followed by a T-duality transformation along $y^1$. The
resulting metric in the IIB theory is:
\beq
ds^2_{10}\,=\,\big[\,H(\rho)\,\big]^{-{1\over 2}}\,
\Big[\,dx^2_{1,2}\,+\,(dy^1)^2\,\Big]\,+\,
\big[\,H(\rho)\,\big]^{{1\over 2}}\,ds^2_6\,\,,
\label{ctdseis}
\eeq
while the dilaton is constant and there is a RR 5-form:
\beq
F^{(5)}=\partial_{\rho}\,\Big[\,H(\rho)\,\Big]^{-1}\,\,
dx^0\wedge dx^1\wedge dx^2\wedge  dy\wedge d\rho\,\,+\,\,
{\rm Hodge\,\, dual}\,\,.
\label{ctdsiete}
\eeq
This solution is precisely the one studied in ref.\cite{pzt2} and
corresponds to a D3--brane located at the tip of the generalized
resolved conifold.

\subsubsection{Smeared D2--D6 wrapped on a 2-cycle}

Another possibility is to reduce along the fiber $\psi$ of the $T^{1,1}$ space.
However, notice that $\psi$ is actually the angle $\tilde \phi$ on the left
invariant 1--forms corresponding to the external $S^3$. That is, in order to
KK reduce along it, a $\psi$--dependent local Lorentzian rotation is
necessary \footnote{We thank Toni Mateos for pointing out this issue to
us.}, say, to go to the canonical vierbein basis $\tilde e^1 = d\tilde\theta$,
$\tilde e^2 = \sin\tilde\theta d\tilde\psi$, $\tilde e^3 = d\tilde\phi +
\cos\tilde\theta d\tilde\psi$, this naturally introducing a functional
dependence on the eleven dimensional Killing spinor \cite{bgm}. This
will render, as we will see, a non--supersymmetric supergravity solution.
This is nothing but the phenomenon of supersymmetry without supersymmetry
first discussed in \cite{dlp}. In order to write the result of the reduction
along $\psi$ (or $\tilde\phi$), let us define the function:
\beq
\Gamma(\rho)\,\equiv\,{\rho^2\over 9}\,\kappa(\rho)\,\,.
\label{ctdocho}
\eeq
Then, the solution of the type IIA theory that one obtains by reducing along
$\psi$ is:
\bear
ds^2_{10}&=&\Bigg[\,{\Gamma(\rho)\over H(\rho)}\,\Bigg]^{{1\over 2}}\,\,
\Big[\,dx^2_{1,2}\,+\,H(\rho)\,\big(\,dy^2_2\,+\,
{d\rho^2\over \kappa(\rho)}\,+\,{\rho^2+6a^2\over 6}\,d\Omega^2_{2,1}\,+\,
{\rho^2\over 6}\,d\Omega^2_{2,2}
\,\,\big)\,\Big]\,\,,\rc\rc
e^{\phi_D}&=&\Big[\,\Gamma(\rho)\,\Big]^{{3\over 4}}\,\,
\Big[\,H(\rho)\,\Big]^{{1\over 4}}\,\,,\rc\rc
F^{(2)}&=&\epsilon^1_{(2)}\,+\,\epsilon^2_{(2)}\,\,,\rc\rc
F^{(4)}&=&\partial_{\rho}\,\Big[\,H(\rho)\,\Big]^{-1}\,\,
dx^0\wedge dx^1\wedge dx^2\wedge  d\rho\,\,,
\label{ctdnueve}
\eear
where $d\Omega^2_{2,a}\,=\,d\theta_a^2\,+\,\sin^2\theta_a d\phi_a^2$ and
$\epsilon^a_{(2)}\,=\,\sin\theta_a d\phi_a\wedge d\theta_a$ for $a=1,2$. This
(non--supersymmetric) solution (of IIA supergravity) corresponds to a system
of (D2-D6)-branes, with the D2-brane extended along $(x^1, x^2)$ and smeared
in $(y^1,y^2)$ and the D6-brane wrapped on a two cycle. Notice that the KK
reduction somehow disentangled the bundle and the resulting ten dimensional
metric exhibits a product of the two-spheres instead of a fibration. This is
characteristic of what has been called supersymmetry without supersymmetry:
the supergravity solution does not display supersymmetry even when it may
be present at the level of full string theory \cite{dlp,bgm,mt}.

\subsubsection{D4--branes}

If we now perform  T-duality transformations along the coordinates $(y^1,
y^2)$, we arrive at a system composed by D4-branes, for which the metric
and dilaton are:
\bear
ds^2_{10}&=&\Bigg[\,{\Gamma(\rho)\over H(\rho)}\,\Bigg]^{{1\over 2}}\,\,
\Big[\,dx^2_{1,2}\,+\,{dy^2_2\over \Gamma(\rho)}\,
+\,H(\rho)\,\big(\,
{d\rho^2\over \kappa(\rho)}\,+\,{\rho^2+6a^2\over 6}\,d\Omega^2_{2,1}\,+\,
{\rho^2\over 6}\,d\Omega^2_{2,2}
\,\,\big)\,\Big]\,\,,\rc\rc
e^{\phi_D}&=&
\Bigg[\,{\Gamma(\rho)\over H(\rho)}\,\Bigg]^{{1\over 4}}\,\,.
\label{ctveinte}
\eear
Moreover, the direct application of the T-duality rules gives the following RR
potentials:
\bear
C^{(3)}&=&\cos\theta_1\,d\phi_1\wedge dy^1\wedge dy^2\,+\,
\cos\theta_2\,d\phi_2\wedge dy^1\wedge dy^2\,\,,\rc\rc
C^{(5)}&=&\big[\,H(\rho)\,\big]^{-1}\,
dx^0\wedge dx^1\wedge dx^2\wedge dy^1\wedge dy^2\,\,.
\label{ctvuno}
\eear
However, since  $C^{(5)}$ is really the potential of
$F^{(6)}\,=\,{}^{*}\,F^{(4)}$, we will only have a four-form  RR field
strength, given by:
\beq
F^{(4)}\,=\,(\,\epsilon^1_{(2)}\,+\,
\epsilon^2_{(2)}\,)\,\wedge dy^1\wedge dy^2\,+\,{k\over 27}\,
\epsilon^1_{(2)}\wedge \epsilon^2_{(2)}\,\,,
\label{ctvdos}
\eeq
where $k$ is the constant appearing in the harmonic function $H(\rho)$.
Again, this solution displays the supersymmetry without supersymmetry
behavior.

\medskip
\section{Summary and Conclusions}
\medskip

In this paper we have studied supergravity solutions corresponding to
D6-branes which wrap two- and three-cycles and have a four-form flux
along the non compact directions of their worldvolume. These solutions are
found first in eight-dimensional gauged supergravity by solving a system of
first-order equations which arises by requiring that the solution be
supersymmetric or, equivalently, by deriving them from  a superpotential in the
corresponding effective lagrangian problem. After uplifting them to eleven
dimensions, our solutions give rise to geometries which are the small
resolution of the conifold (for D6-branes wrapping a two cycle) or a manifold
of $G_2$ holonomy (in the case of a D6-brane wrapping an $S^3$ in $T^*S^3$),
with the corresponding warp factors included. The latter are the effect of the
four-form flux on the metric, a fact which we have checked in general in
appendix C. These configurations can be interpreted as smeared M2-branes on the
tip of a (resolved) cone. By performing different Kaluza-Klein reductions and
T-dualities we have obtained several solutions corresponding to D2, D2-D6, D3
and D4 systems and, in some cases, we have discussed the corresponding field
theory duals in 2+1 dimensions. Some of them display supersymmetry without
supersymmetry.

Let us finally point out some directions which would be worth to explore 
in future. First of all, it is clear that it would be desirable to have a
better understanding of the field theory duals to the supergravity solutions
studied here. Moreover, it would be also interesting to look at solutions which
also have non-vanishing Neveu-Schwarz fluxes, as those studied in ref.
\cite{bgm}. Another interesting problem to look at is the generation, in the
framework of D=8 gauged supergravity, of solutions with non-vanishing
components of the 4-form along the compact directions of the special holonomy
manifold. The corresponding field theory duals would be three-dimensional gauge
theories with  Chern-Simons terms. We could also try to generalize our ansatzs
(\ref{uno}) and (\ref{stocho}) for the metric to the case in which the D2-brane
is localized in the $y$ direction. We hope to report on these issues in a near
future.

\medskip
\section*{Acknowledgments}
\medskip
We are grateful to Joaquim Gomis, Juan Maldacena, Toni Mateos, Carlos
N\'u\~nez and Martin Schvellinger for highly valuable comments. JDE wishes
to thank the members of the
Theory Group in Santiago de Compostela for their kind hospitality.
This work was supported in part by DGICYT under grant PB96-0960, by CICYT under
grant AEN99-0589-CO2-02, by Xunta de Galicia under grant PGIDT00-PXI-20609,
by Fundaci\'on Antorchas and by Funda\c c\~ao para a Ci\^encia e a Tecnologia
under grants POCTI/1999/MAT/33943 and SFRH/BPD/7185/2001.

\medskip
\section*{Appendices}
\medskip

\renewcommand{\theequation}{\rm{A}.\arabic{equation}}
\setcounter{equation}{0}
\medskip
\appendix
\section{D=8 gauged supergravity}
\medskip
The maximal eight dimensional gauged supergravity was constructed by Salam and
Sezgin in ref. \cite{ss} by means of a dimensional reduction of D=11
supergravity
on a
$SU(2)$ group manifold \cite{ssch}. The bosonic field content of this theory
can be consistently truncated to include the metric $g_{\mu\nu}$, a dilatonic
scalar $\phi$, five scalars parametrized by a $3\times 3$
unimodular matrix $L_{\alpha}^i$ which lives in the coset
$SL(3,\RR)/SO(3)$, an $SU(2)$ gauge potential $A_{\mu}^i$ and a three-form
potential $B$. The kinetic energy of the coset scalars $L_{\alpha}^i$ is given
in terms of the symmetric traceless  matrix $P_{\mu\,ij}$ defined by means of
the expression:
\beq
P_{\mu\,ij}+Q_{\mu\,ij}\,\,=\,L_i^{\alpha}\,
(\,\partial_{\mu}\,\delta_{\alpha}^{\beta}\,-\,
g\,\epsilon_{\alpha\beta\gamma}\,A_{\mu}^{\gamma}\,)\,L_{\beta j}\,\,,
\label{apauno}
\eeq
where $Q_{\mu\,ij}$ is, by definition, the antisymmetric part of the right-hand
side of eq. (\ref{apauno}). Furthermore, the potential energy of the coset
scalars is written in terms of the so-called $T$-tensor, $T^{ij}$, and of its
trace, $T$, defined as:
\beq
T^{ij}\,=\,L^i_{\alpha}\,L^j_{\beta}\,\delta^{\alpha\beta}\,\,,
\,\,\,\,\,\,\,\,\,\,\,\,\,\,\,\,\,\,\,\,\,\,\,\,
T\,=\,\delta_{ij}\,T^{ij}\,\,.
\label{apados}
\eeq
If $F_{\mu\nu}^{i}$ is the field strength of the $SU(2)$ gauge field
$A_{\mu}^i$ and if $G_{\mu\nu\rho\sigma}$ denotes the components of $dB$, the
bosonic lagrangian for this truncation of D=8 gauged supergravity is:
\bear
{\cal L}&=&\sqrt{-g_{(8)}}\,\,\Big[\,
{1\over 4}\,R\,-\,{1\over 4}\,e^{2\phi}\,
F_{\mu\nu}^{i}\,F^{\mu\nu\,\,i}\,-\,{1\over 4}\,
P_{\mu\,ij}\,P^{\mu\,ij}\,-\,{1\over 2}\,
\partial_{\mu}\,\phi\partial^{\mu}\,\phi\,-\cr\cr
&&-{g^2\over 16}\,e^{-2\phi}\,(\,T_{ij}\,T^{ij}\,-\,{1\over 2}T^2\,)\,-\,
{1\over 48}\,e^{2\phi}\,G_{\mu\nu\rho\sigma}\,
G^{\mu\nu\rho\sigma}\,\,\Big]\,\,,
\label{apatres}
\eear
and the corresponding equations of motion are:
\bear
R_{\mu\nu}&=&P_{\mu\,ij}P_{\nu}^{ij}\,+\,
2\partial_{\mu}\phi\partial_{\nu}\phi\,+\,
2e^{2\phi}\, F_{\mu\lambda}^{i}\,F_{\nu}^{\lambda\, i}\,-\,
{1\over 3}\, g_{\mu\nu}\nabla^2\,\phi\,+\rc\rc
&&+\,{1\over 3}\,e^{2\phi}\, \big(\,
G_{\mu\lambda\tau\sigma}\,G_{\nu}^{\,
\,\lambda\tau\sigma}\,-\,{1\over 12}\,g_{\mu\nu}\,
G_{\rho\lambda\tau\sigma}\,G^{\rho\lambda\tau\sigma}\,\big)\,\,,\rc\rc
D_{\mu}\,\Big(\, \sqrt{-g_{(8)}}\,P^{\mu ij}\,\Big)&=&
-{2\over 3}\, \nabla^2\, \phi\, \delta^{ij}\,
+\, e^{2\phi}\, F_{\mu\nu}^iF^{\mu\nu\, j}
\,+\, {1\over 36}\, e^{2\phi}\,
G_{\mu\nu\tau\sigma}\,G^{\mu\nu\tau\sigma}\,\,\delta^{ij}\,+\rc\rc
&&+\, {1\over 2}\, g^2\, e^{-2\phi}\,
\big[\,T^{ik}\,T^{jk}\,-\,{1\over 2}\,T\,T^{ij}\,-\,{1\over 2}\,\delta^{ij}\,
(\,T^{kl}T^{kl}\,-\,{1\over 2}\,T^2\,)\,\big]\,\,,\rc\rc
D_{\mu}\,\Big(\, \sqrt{-g_{(8)}}\,
e^{2\phi}\, F^{\mu\nu\,i}\,\Big)&=&-e^{2\phi}\,P_{\mu}^{ij}\,
F^{\mu\nu\,j}\,-\,g\,g^{\mu\nu}\,\epsilon_{ijk}\,P_{\mu}^{jl}\,
T^{kl}\,\,,\rc\rc
D_{\mu}\,\Big(\, \sqrt{-g_{(8)}}\,\,e^{2\phi}\,
G^{\mu\nu\tau\sigma}\,\big)\,&=&0\,\,.
\label{apacuatro}
\eear
Given a solution of the equations (\ref{apacuatro}) one can write a solution of
the D=11 theory by reverting the Salam-Sezgin reduction ansatz. For the eleven
dimensional metric, the corresponding uplifting formula is:
\beq
ds^2_{11}\,=\,e^{-{2\over 3}\,\phi}\,ds^2_8\,+\,4\,e^{{4\over 3}\phi}\,
(\,A^i\,+\,{1\over 2}\,L^i\,)^2\,\,,
\label{apacinco}
\eeq
where $L^i$ is defined as:
\beq
L^i\,=\,{2\over g}\,\tilde w^{\alpha}\,L_{\alpha}^i\,\,,
\label{apaseis}
\eeq
with $\tilde w^{i}$ being left invariant forms on the $SU(2)$ group
manifold and $g$ is the $SU(2)$ gauge coupling constant. The relation
between the eleven and eight dimensional four-forms has been given
in eq.(\ref{cseis}).

\medskip
\renewcommand{\theequation}{\rm{B}.\arabic{equation}}
\setcounter{equation}{0}
\section{Effective lagrangian with 4-form}
\medskip
In this appendix we explain how to find effective lagrangians for a given
ansatz for the eight dimensional fields when the four-form $G$ is non-zero. Let
us imagine that we substitute our ansatz for the metric and gauge field
$A_{\mu}^i$ in the Salam-Sezgin lagrangian (\ref{apatres}) and let us
denote by $f_i$ the different functions $f,h, \alpha, \cdots$ of the
ansatz (including the dilaton and other scalar fields).
As the four-form field has a radial component, we can represent it as $B'$,
where $B$ is a potential and the prime denotes radial derivative. After
integrating by parts to eliminate the second derivatives, the resulting
lagrangian will be of the type:
\beq
{\cal L}\,=\,\tilde {\cal L}(\,f_i, f_i'\,)\,+\,a(\,f_i\,)\,
(\,B'\,)^2\,\,,
\label{apbuno}
\eeq
where $a(\,f_i\,)$ does not depend on the derivatives of the $f_i$'s.
The equations of motion for ${\cal L}$  are:
\bear
{d\over dr}\,{\partial \tilde {\cal L}\over \partial f_i'}&=&
{\partial \tilde {\cal L}\over \partial f_i}\,+\,(\,B'\,)^2\,
{\partial a\over \partial f_i}\,\,,\rc\rc
{d\over dr}\,\Big[\,a\,B'\,\Big]&=&0\,\,,
\label{apbdos}
\eear
which, together with the corresponding zero energy condition, are
equivalent to
(\ref{apacuatro}). Integrating the equation for $B$ we get:
\beq
B'\,=\,{\Lambda\over a(\,f_i\,)}\,\,,
\label{apbtres}
\eeq
where $\Lambda$ is a constant. This is precisely our ansatz for $G$ in
eqs.(\ref{dos}) and (\ref{stnueve}). Substituting the value of $B'$ given
in eq.(\ref{apbtres}) in the equation for the $f_i$'s, one gets:
\beq
{d\over dr}\,{\partial \tilde {\cal L}\over \partial f_i'}=
{\partial \tilde {\cal L}\over \partial f_i}\,+\,
{\Lambda^2\over a^2}\,{\partial a\over \partial f_i}\,=\,
{\partial \over \partial f_i}\,\Big(\,
\tilde {\cal L}\,-\,{\Lambda^2\over a}\,\Big)\,\,,
\label{apbcuatro}
\eeq
and, therefore,  the effective lagrangian for the $f_i$'s is:
\beq
{\cal L}_{eff}\,=\,
\tilde {\cal L}(\,f_i, f_i'\,)\,-\,
{\Lambda^2\over a(\,f_i\,)}\,\,.
\label{apbcinco}
\eeq
Indeed, the Euler-Lagrange equations for ${\cal L}_{eff}$ are precisely
(\ref{apbcuatro}). Notice the change of sign in the last term of
${\cal L}_{eff}$ as compared with the corresponding one in ${\cal L}$.
This sign flip has been taken into account is eqs.(\ref{quince}) and
(\ref{odos}) and is crucial to find the superpotentials. Equivalently, one can
obtain ${\cal L}_{eff}$ by eliminating the cyclic coordinate $B$ by
constructing
the Routhian ${\cal R}$ as:
\beq
{\cal R}\,=\,{\cal L}\,-\,B'\,{\partial {\cal L}\over \partial B'}\,\,.
\label{apbseis}
\eeq
Clearly ${\cal R}={\cal L}_{eff}$.

\medskip
\renewcommand{\theequation}{\rm{C}.\arabic{equation}}
\setcounter{equation}{0}
\section{General dependence  of the metric on the 4-form}
\medskip

In sections 2 and 3 we have concluded that the effect of the four-form
on the metric, as compared to the one obtained when the four-form is set to
zero,  is just the introduction of some warp factors. In this appendix we will
prove that the validity of this result goes beyond the particular cases
studied in the main text and, under some conditions, it holds in general.

Let us suppose that we adopt the following ansatz for the  eight-dimensional
metric and 4-form:
\bear
ds_8^2\,&=&\,e^{2f}\,dx_{1,2}^2\,+\,\sum_{i=1}^{4}\,
e^{2h_i}\,\big(\,E^i\,)^2\,+\,dr^2\,\,,\rc\rc
G_{\underline{x^0x^1x^2r}}&=&\Lambda\,
e^{-\sum h_i\,-\,2\phi}\,\equiv\,\Lambda\,e^{-\phi}\,
\xi(\,\phi, h_i\,)\,\,,
\label{apcuno}
\eear
where $E^i$ are some vierbiens, which we will assume to be independent of
the radial coordinate $r$, $\Lambda$ is a constant and we have defined the
function $\xi(\,\phi, h_i\,)$. The equation of motion  for $G$ is satisfied
when $G$ has the form given in eq.(\ref{apcuno}). All the dependence on $r$ is
included in the functions $f$, $h_i$ and $\phi$. We will assume that we
have also some scalar fields $\lambda_i$. These functions satisfy certain
first-order BPS equations of the type:
\bear
{d\over dr}\,f&=&\Gamma_f\,(\,\phi,h_i, \lambda_i)\,+\,
{\Lambda\over 2}\,\,\xi(\,\phi, h_i\,)\,\,,\rc\rc
{d\over dr}\,h_i&=&\Gamma_{h_i}\,(\,\phi,h_i, \lambda_i)\,-\,
{\Lambda\over 2}\,\,\xi(\,\phi, h_i\,)\,\,,\rc\rc
{d\over dr}\,\phi&=&\Gamma_{\phi}\,(\,\phi,h_i, \lambda_i)\,-\,
{\Lambda\over 2}\,\,\xi(\,\phi, h_i\,)\,\,,\rc\rc
{d\over dr}\,\lambda_i&=&\Gamma_{\lambda_i}\,(\,\phi,h_i, \lambda_i)\,\,,
\label{apcdos}
\eear
where the functions $\Gamma$ of the right-hand side depend on the particular
case we are considering. The only property we will need of these functions is
that they satisfy the following homogeneity condition:
\beq
\Gamma(\,\phi\,+\,\gamma\,,\,h_i+\,\gamma\,, \lambda_i\,)\,=\,
e^{-\gamma}\,\Gamma(\,\phi\,,\,h_i\,, \lambda_i\,)\,\,,
\label{apctres}
\eeq
where $\gamma$ is an arbitrary function. In all the cases studied here and in
refs. \cite{gns, hs2} the $\Gamma$'s satisfy (\ref{apctres}). On the other
hand, from the definition of $\xi(\,\phi\,,\,h_i\,)$ one has:
\beq
\xi(\,\phi\,+\,\gamma\,,\,h_i+\,\gamma\,)\,=\,
e^{-5\gamma}\,\xi(\,\phi\,,\,h_i\,)\,\,.
\label{apccuatro}
\eeq

Let us now consider a function $\eta$ such that solves the following
differential equation:
\beq
{d\eta\over dr}\,=\,-{\Lambda\over 2}\,\,\xi(\,\phi, h_i\,)\,\,,
\label{apccinco}
\eeq
and let us define the functions:
\beq
\tilde f\,=\,f\,+\,\eta\,\,,
\,\,\,\,\,\,\,\,\,\,\,\,\,\,
\tilde h_i\,=\,h_i\,-\,\eta\,\,,
\,\,\,\,\,\,\,\,\,\,\,\,\,\,
\tilde \phi\,=\,\phi\,-\,\eta\,\,.
\label{apcseis}
\eeq
If we now introduce a new radial variable $\tilde r$ such that:
\beq
{d  r\over d\tilde r}\,=\,e^{\eta}\,\,,
\label{apcsiete}
\eeq
then,  it is straightforward to prove that $\tilde f\,$, $\tilde h_i\,$ and
$\tilde \phi$ and $\lambda$ satisfy the following differential equations:
\bear
{d\over d\tilde r}\,\tilde f&=&
\Gamma_f\,(\,\tilde\phi,\tilde h_i, \lambda_i)\,\,,\rc\rc
{d\over d\tilde r}\,\tilde h_i&=&
\Gamma_{h_i}\,(\,\tilde \phi,\tilde h_i,\lambda_i)\,\,,\rc\rc
{d\over d\tilde r}\,\tilde\phi&=&
\Gamma_{\phi}\,(\,\tilde\phi,\tilde h_i, \lambda_i)\,\,,\rc\rc
{d\over d\tilde r}\,\lambda_i&=&
\Gamma_{\lambda_i}\,(\,\tilde \phi,\tilde h_i, \lambda_i)\,\,,
\label{apcocho}
\eear
which are   the same  as in those for the same system without the 4-form.
Moreover, if we define the function $H$ as:
\beq
H\,\equiv\,e^{4\eta}\,\,,
\label{apcnueve}
\eeq
then, the uplifted metric is:
\bear
ds^2_{11}\,&=&\,H^{-{2\over 3}}\,
e^{2\tilde f\,-\,{2\over  3}\,\tilde\phi}\,dx^2_{1,2}\,+\,\rc\rc
&&+\,H^{{1\over 3}}\,\Big[\,\sum_i\,e^{2\tilde h_i\,-\,{2\over  3}\tilde\phi}
\big(\,E^i\,)^2\,+\,e^{-{2\over  3}\tilde\phi}\,d\tilde r^2
\,+\,4\,e^{{4\over  3}\tilde\phi}
\,\Big(\,A_i\,+\,{1\over 2}\,L_i\,\Big)^2\,\Big]\,\,.
\label{apcdiez}
\eear

It is clear from eq. (\ref{apcdiez}) that
the effect of the 4-form on the metric is the introduction of some powers
of $H$ which distinguish among  the directions parallel and orthogonal to the
form. Moreover, it is easy to verify from the equation satisfied by $\eta$ that
the harmonic function $H$ satisfies:
\beq
{dH\over d\tilde r}\,=\,-2\Lambda\,\,
\xi(\,\tilde \phi\,, \,\tilde h^i\,)\,=\,-2\Lambda
e^{-\sum \tilde h_i\,-\,\tilde\phi}\
\,\,,
\label{apconce}
\eeq
and, thus, if we know the solution without form, we can integrate the
right-hand side of the last equation and find the expression of $H$. Notice
that when $\Lambda=0$ we can take $H={\rm constant}$. In this case the
components of the metric parallel to the 4-form are constant provided
that $\tilde\phi=3\tilde f$ solves eq. (\ref{apcocho}), which can only happen
if $\Gamma_\phi=3\Gamma_f$. This condition holds for all the systems studied
here and in refs. \cite{gns, hs2}. Moreover, if $\tilde\phi=3\tilde f$
one can verify that the uplifted 4-form is such that
$F_{x^0x^1x^2\tilde r}=\,\partial_{\tilde r}\,(\,H^{-1}\,)$.

As a illustration of the general formalism we have developed above, let us
consider the case of a flat D6-brane with flux. In this situation there are no
scalar fields $\lambda$ excited and the ansatz for the metric is \cite{en}:
\beq
ds^2_8\,=\,e^{2 f}\,dx_{1,2}^2\,+\,e^{2 h}\,dy_4^2\,+\,d r^2\,\,.
\label{apcdoce}
\eeq
The functions $\Gamma$ appearing in the first-order system (\ref{apcdos}) are
$\Gamma_f=\Gamma_h={\Gamma_{\phi}\over 3}={g\over 8}\,e^{-\phi}$. If we change
to a new variable $t$ such that  $d\tilde r\,=\,e^{-\tilde\phi}\,dt$, we can
write the solution of the system (\ref{apcocho}) as
$\tilde f=\tilde h={\tilde \phi\over 3}={g\over 8}\,t$. Moreover, for the case
at hand $\xi(\,\tilde \phi\,, \,\tilde h\,)\,=\,e^{-4\tilde h-\tilde\phi}$ and,
by plugging this result in eq. (\ref{apconce}) and taking $g=1$, we get
that the
harmonic function is:
\beq
H\,=\,-2\Lambda\,\int e^{-4\tilde h-\tilde\phi} d\tilde r\,=\,
-2\Lambda\,\int e^{-4\tilde h} dt\,=\,1\,+\,4\Lambda\,e^{-{t\over 2}}\,\,,
\label{apctrece}
\eeq
where we have fixed the integration constant to recover the solution with
$\Lambda=0$ at $t\rightarrow\infty$.  The eleven dimensional metric is readily
obtained from the uplifting formula (\ref{apacinco}). Since there are no SU(2)
gauge fields excited in this flat case \cite{en}, we get:
\beq
ds^2_{11}\,=\,H^{-{2\over 3}}\,dx^2_{1,2}\,+\,H^{{1\over 3}}          
\,\,\Big(\,dy_4^2\,+\,e^{{t\over 2}}\,(dt^2\,+\,16\,d\Omega_3^2)\,\Big)\,\,.
\label{apccatorce}
\eeq
Introducing a new variable
$\rho$ as $\rho\,=\,{4\over {\sqrt N}}\,e^{{t\over 4}}$, the metric
(\ref{apccatorce}) can be put in the form:
\beq
ds^2_{11}\,=\,\Big[\,H(\rho)\,\Big]^{-{2\over 3}}\,dx^2_{1,2}\,+\,
\Big[\,H(\rho)\,\Big]^{{1\over 3}}
\,\,\Big(\,dy_4^2\,+\,
N(d\rho^2\,+\,\rho^2\,d\Omega_3^2\,)\,\Big)\,\,,
\label{apcquince}
\eeq
where $H(\rho)$ is given by:
\beq
H(\rho)\,=\,1\,+\,{64\Lambda\over N}\,{1\over \rho^2}\,\,.
\label{apcdseis}
\eeq
Notice that, as pointed out in the main text,  the harmonic function of the
D2-brane $H(\rho)$ appearing  in the metric (\ref{apcquince}) is not in its
near horizon limit. Actually, if one drops the $1$ on the right-hand side of
eq. (\ref{apcdseis}), one can check that (\ref{apccatorce}) coincides with the
metric of the standard near horizon D2-D6 intersection.


\begin{thebibliography}{99}

\bibitem{bvs} M.~Bershadsky, C.~Vafa and V.~Sadov, ``D-Branes and Topological
Field Theories'', Nucl.\ Phys.\ B {\bf 463} (1996) 420, {\rm hep-th/9511222}.

\bibitem{jm} J.~M.~Maldacena, ``The large $N$ limit of superconformal field
theories and supergravity'', Adv.\ Theor.\ Math.\ Phys.\  {\bf 2} (1998) 231,
{\rm hep-th/9711200}.

\bibitem{mn} J.~M.~Maldacena and C.~N\'u\~nez, ``Supergravity description of
field theories on curved manifolds and a no go theorem'', {\rm hep-th/0007018}.
~``Towards the large N limit of pure ${\cal N} = 1$ super Yang Mills'', Phys.\
Rev.\ Lett.\ {\bf 86} (2001) 588 {\rm hep-th/0008001}.

\bibitem{ks} I.~R.~Klebanov and M.~J.~Strassler, ``Supergravity and a
confining gauge theory: Duality cascades and (chi)SB-resolution of naked
singularities'', JHEP{\bf 0008} (2000) 052, {\rm hep-th/0007191}.

\bibitem{imsy} N.~Itzhaki, J.~M.~Maldacena, J.~Sonnenschein and
S.~Yankielowicz, ``Supergravity and the large N limit of theories with
sixteen supercharges'', Phys.\ Rev.\ D {\bf 58} (1998) 046004, {\rm
hep-th/9802042}.

\bibitem{gauged} B.~S.~Acharya, J.~P.~Gauntlett and N.~Kim, ``Fivebranes
wrapped on associative three-cycles'', Phys.\ Rev.\ D {\bf 63} (2001)
106003, {\rm hep-th/0011190}. ~H.~Nieder and Y.~Oz, ``Supergravity and
D-branes wrapping special Lagrangian cycles'', JHEP{\bf 0103} (2001)
008, {\rm hep-th/0011288}. J.~P.~Gauntlett, N.~Kim and D.~Waldram,
``M-fivebranes wrapped on supersymmetric cycles'', Phys.\ Rev.\ D {\bf
63} (2001) 126001, {\rm hep-th/0012195}. ~C.~N\'u\~nez, I.~Y.~Park,
M.~Schvellinger and T.~A.~Tran, ``Supergravity duals of gauge theories
from F(4) gauged supergravity in six dimensions'', JHEP {\bf 0104} (2001)
025, {\rm hep-th/0103080}.

\bibitem{en} J.D.~Edelstein and C.~N\'u\~nez, ``D6 branes and M-theory
geometrical transitions from gauged  supergravity'',
{\sl \jhep} {\bf 0104}, (2001) 028, {\rm hep-th/0003037}.

\bibitem{st} M.~Schvellinger and T.~A.~Tran, ``Supergravity duals of gauge
field theories from SU(2) x U(1) gauged  supergravity in five dimensions'',
JHEP {\bf 0106} (2001) 025, {\rm hep-th/0105019}. ~J.~M.~Maldacena and
H.~Nastase, ``The supergravity dual of a theory with dynamical
supersymmetry breaking'', JHEP {\bf 0109} (2001) 024, {\rm hep-th/0105049}.
~J.~P.~Gauntlett, N.~Kim, S.~Pakis and D.~Waldram, ``Membranes wrapped on
holomorphic curves'', Phys.\ Rev.\ D {\bf 65} (2002) 026003, {\rm
hep-th/0105250}. ~R.~Hernandez, ``Branes wrapped on coassociative cycles",
{\sl \pl} {\bf B521}(2001) 371, {\rm hep-th/0106055}. ~J.~P.~Gauntlett,
N.~Kim, D.~Martelli and D.~Waldram, ``Wrapped fivebranes and N = 2 super
Yang-Mills theory'', Phys.\ Rev.\ D {\bf 64} (2001) 106008, {\rm
hep-th/0106117}. ~F.~Bigazzi, A.~L.~Cotrone and A.~Zaffaroni, ``N = 2 gauge
theories from wrapped five-branes'', Phys.\ Lett.\ B {\bf 519} (2001) 269,
{\rm hep-th/0106160}. ~J.~Gomis and T.~Mateos, ``D6 branes wrapping Kaehler
four-cycles'', Phys.\ Lett.\ B {\bf 524} (2002) 170, {\rm hep-th/0108080}.
~J.~P.~Gauntlett and N.~Kim, ``M-fivebranes wrapped on supersymmetric
cycles. II'', Phys.\ Rev.\ D {\bf 65} (2002) 086003, {\rm hep-th/0109039}.

\bibitem{gr} J.~Gomis and J.~G.~Russo, ``D = 2+1 N = 2 Yang-Mills theory
from wrapped branes'', JHEP {\bf 0110} (2001) 028, {\rm hep-th/0109177}.

\bibitem{gkmw} J.~P.~Gauntlett, N.~Kim, D.~Martelli and D.~Waldram,
``Fivebranes wrapped on SLAG three-cycles and related geometry'',
JHEP {\bf 0111} (2001) 018, {\rm hep-th/0110034}. ~J.~Gomis, ``On SUSY
breaking and chiSB from string duals'', Nucl.\ Phys.\ B {\bf 624} (2002)
181, {\rm hep-th/0111060}. ~G.~Curio, B.~Kors and D.~Lust, ``Fluxes and
branes in type II vacua and M-theory geometry with G(2) and Spin(7)
holonomy'', Nucl.\ Phys.\ B {\bf 636} (2002) 197, {\rm hep-th/0111165}.
~P.~Di Vecchia, H.~Enger, E.~Imeroni and E.~Lozano-Tellechea, ``Gauge
theories from wrapped and fractional branes'', Nucl.\ Phys.\ B {\bf 631}
(2002) 95, {\rm hep-th/0112126}. R. Apreda, F. Bigazzi, A. L. Cotrone, M.
Petrini and A. Zaffaroni, ``Some comments on N=1 gauge theories from wrapped
branes'', Phys.\ Lett.\ B {\bf 536} (2002) 161, {\rm hep-th/0112236}.
~R.~Hernandez and K.~Sfetsos, ``An
eight-dimensional approach to G(2) manifolds'', Phys.\ Lett.\ B {\bf 536}
(2002) 294, {\rm hep-th/0202135}. ~J.~P.~Gauntlett, N.~Kim, S.~Pakis and
D.~Waldram, ``M-theory solutions with AdS factors'', {\rm hep-th/0202184}.
~P.~Di Vecchia, A.~Lerda and P.~Merlatti, ``N = 1 and N = 2 super
Yang-Mills theories from wrapped branes'', {\rm hep-th/0205204}.
~M.~Naka, ``Various wrapped branes from gauged supergravities'',
{\rm hep-th/0206141}.

\bibitem{ssch} J.~Scherk and J.~H.~Schwarz, ``How To Get Masses From Extra
Dimensions'', {\sl \np} {\bf B153} (1979) 61.

\bibitem{ps} O.~Pelc and R.~Siebelink, ``The D2-D6 system and a fibered AdS
geometry'', Nucl.\ Phys.\ B {\bf 558} (1999) 127, {\rm hep-th/9902045}.

\bibitem{lo} A.~Loewy and Y.~Oz, ``Branes in special holonomy backgrounds'',
{\rm hep-th/0203092}.

\bibitem{gns} U.~Gursoy, C.~N\'u\~nez and M.~Schvellinger, ``RG flows from
Spin(7), CY 4-fold and HK manifolds to AdS, Penrose  limits and pp waves'',
{\rm hep-th/0203124}.

\bibitem{dlp} M.~J.~Duff, H.~Lu and C.~N.~Pope, ``Supersymmetry without
supersymmetry'', Phys.\ Lett.\ B {\bf 409} (1997) 136, {\rm hep-th/9704186}.

\bibitem{ss} A.~Salam and E.~Sezgin, ``d=8 Supergravity'',
{\sl \np} {\bf B258} (1985) 284.

\bibitem{hs2} R.~Hernandez and K.~Sfetsos, ``Branes with fluxes wrapped on
spheres'', {\rm hep-th/0205099}.

\bibitem{bs} R.~Bryant and S.~Salamon, ``On the construction of some
complete metrics with exceptional holonomy'', Duke\ Math.\ J.\ {\bf 58}
(1989) 829.

\bibitem{gpp} G.~W.~Gibbons, D.~N.~Page and C.~N.~Pope, ``Einstein
metrics on $S^3$, $R^3$ And $R^4$ Bundles'', Commun.\ Math.\ Phys.\
{\bf 127} (1990) 529.

\bibitem{bb} K.~Becker and M.~Becker, ``M-Theory on Eight-Manifolds''
Nucl.\ Phys.\ B {\bf 477} (1996) 155, {\rm hep-th/9605053}.

\bibitem{local} R.~R.~Khuri, ``Remark On String Solitons'', Phys.\ Rev.\ D
{\bf 48} (1993) 2947, {\rm hep-th/9305143}. ~J.~P.~Gauntlett, D.~A.~Kastor
and J.~Traschen, ``Overlapping Branes in M-Theory'', Nucl.\ Phys.\ B {\bf
478} (1996) 544, {\rm hep-th/9604179}. ~J.~D.~Edelstein, L.~Tataru and
R.~Tatar, ``Rules for localized overlappings and intersections of p-branes'',
JHEP {\bf 9806} (1998) 003, {\rm hep-th/9801049}. ~N.~Itzhaki, A.~A.~Tseytlin
and S.~Yankielowicz, ``Supergravity solutions for branes localized within
branes'', Phys.\ Lett.\ B {\bf 432} (1998) 298, {\rm hep-th/9803103}.
~A.~Hashimoto, ``Supergravity solutions for localized intersections of
branes'', JHEP {\bf 9901} (1999) 018, {\rm hep-th/9812159}. ~D.~Youm,
``Localized intersecting BPS branes'', {\rm hep-th/9902208}. ~A.~Fayyazuddin
and D.~J.~Smith, ``Localized intersections of M5-branes and four-dimensional
superconformal field theories'', JHEP {\bf 9904} (1999) 030, {\rm
hep-th/9902210}. ~A.~Loewy, ``Semi localized brane intersections in SUGRA'',
Phys.\ Lett.\ B {\bf 463} (1999) 41, {\rm hep-th/9903038}. ~A.~Gomberoff,
D.~Kastor, D.~Marolf and J.~Traschen, ``Fully localized brane intersections:
The plot thickens'', Phys.\ Rev.\ D {\bf 61} (2000) 024012, {\rm
hep-th/9905094}. ~S.~A.~Cherkis, ``Supergravity solution for M5-brane
intersection'', {\rm hep-th/9906203}.

\bibitem{cghp} M.~Cvetic,  G. W. ~Gibbons, H. Lu and C. N. Pope,
``Supersymmetric non-singular fractional D2-branes and NS-NS
2-branes'', {\rm hep-th/0101096}.

\bibitem{kw} I.~R.~Klebanov and E.~Witten, ``Superconformal field theory on
threebranes at a Calabi-Yau  singularity'', Nucl.\ Phys.\ B {\bf 536} (1998)
199, {\rm hep-th/9807080}.

\bibitem{amv} M.~Atiyah, J.~M.~Maldacena and C.~Vafa, ``An M-theory flop as
a large N duality'', J.\ Math.\ Phys.\  {\bf 42} (2001) 3209, {\rm
hep-th/0011256}.

\bibitem{clp} M.~Cvetic, H.~Lu and C.~N.~Pope
``Consistent warped-space Kaluza-Klein reductions, half-maximal gauged
supergravities and $\CC\IP^n$ constructions'', {\sl \np} {\bf B597} (2001) 172,
{\rm hep-th/0007109}.

\bibitem{cdlo}P.~Candelas and X.~C.~de la Ossa, ``Comments On Conifolds'',
Nucl.\ Phys.\ B {\bf 342} (1990) 246.

\bibitem{pzt1} L.~A.~Pando Zayas and A.~A.~Tseytlin, ``3-branes on resolved
conifold'', {\sl \jhep} {\bf 0008} (2000) 028,
{\rm hep-th/0010088}.

\bibitem{pt} G.~Papadopoulos and A.~A.~Tseytlin, ``Complex geometry of
conifolds and 5-brane wrapped on 2-sphere'', Class.\ Quant.\ Grav.\
{\bf 18} (2001) 1333, {\rm hep-th/0012034}.

\bibitem{pzt2} L.~A.~Pando Zayas and A.~A.~Tseytlin, ``3-branes on spaces with
R x S(2) x S(3) topology'', Phys.\ Rev.\ D {\bf 63} (2001) 086006,
{\rm hep-th/0101043}.

\bibitem{bgm} J.~Brugues, J.~Gomis, T.~Mateos and T.~Ramirez,
``Supergravity duals of noncommutative wrapped D6 branes and
supersymmetry without supersymmetry'', {\rm hep-th/0207091}.

\bibitem{mt} R.~Minasian and D.~Tsimpis, ``Hopf reductions, fluxes and
branes'', Nucl.\ Phys.\ B {\bf 613} (2001) 127, {\rm hep-th/0106266}.

\end{thebibliography}
\end{document}